\documentstyle[11pt,amsfonts,amssymb,epsfig]{article}
\begin{document}

\title{{\large\bf Reanalysis of \boldmath{$e^{\pm} p$} Elastic Scattering Data
in Terms of Proton Electromagnetic Formfactors}}
\author{V.~Ezhela\footnote{ezhela@mx.ihep.su},  B.~Polishchuk\footnote{polishchuk@mx.ihep.su} \\
{\it COMPAS group, Institute for High Energy Physics,} \\
{\it 142284, Protvino, Russia}}
\maketitle
\begin{abstract}
      
 We have made the comparison of three parametrizations of the proton
electromagnetic form factors in the space-like region
using the largest data set on $e^{\pm} p$ elastic scattering ever used. 
All models have a correct analytical
structure with a minimal set of well known singularities.
For the first time an effective two-pion intermediate coulombic bound 
state (pionium) was included into GVMD for nucleon formfactors.
For the proton charge radius an estimate derived from our best fits  is as
follows:
$$ r_E^p [\rm fm] = 0.897 \pm 0.002(exp) \pm 0.001(norm) \pm 0.003(models).$$
The traditional ``model independent'' fit by quadratic polynomial gives, in 
our estimation technology, a much cruder estimate
$$ r^p_{E,polinomial} [\rm fm] = 0.887 \pm 0.006(exp) \pm 0.096(norm).$$
Both estimates are closer to the value estimated from the recent hydrogen Lamb 
shift measurements.
     
\end{abstract}


\section*{Introduction}

Since the discovery of the deviation of differential cross sections for $e p$ 
elastic scattering \cite{hofs} from that of calculated for point-like 
nucleons, numerous efforts were devoted to 
measurement of the ``spatial extention'' of 
nucleons and the corresponding electromagnetic form factors (FFs). 
There are more than 50
experimental reports with data on $e^{\pm} p$ elastic scattering, starting from a
pioneering study of Hofstadter and co-workers in 1953-1956, up to recently 
reported new experimental data of Andivahis et al. in 1994 \cite{Andivahis}
(see Table~5.1).

Many attempts were made to model charge and magnetic moment distributions in 
the nucleons, numerous phenomenological parametrizations of (FFs)
were proposed. The majority of proposed parametrizations is based on 
various modifications of generalized vector mesons dominance (GVMD) model
guided by the requirements from general principles of quantum field theories 
and the modest amount is motivated by QCD features. The common features of the 
parameterizations are the requirements of providing reasonable values of the rms
 ``charge'' and ``magnetic'' radii of the nucleon ($t \to 0$ asymptotic ), 
as well as reproducing QCD-like behaviour at large $|t|$.

Unfortunately, it is still impossible to derive the FFs behaviour,
even in the leading asymptotic terms, from the ``first principles'' and any 
extraction of derived physical parameters (such as radii and constants in an
asymptotic behaviour at large $|t|$) would be ``model dependent.'' 

 On the other hand, the precise knowledge of the nucleon FFs with as 
accurate as possible errors treatment is needed in a number of phenomenological 
applications and even in precise tests of well established theories.
The analyses of charged hadrons elastic scattering in the Coulomb-nuclear 
interference regions and of the data on Lamb shifts in atomic hydrogen are 
the well known examples of such applications.

It is often argued that the extraction of FFs from the experimental 
data is an ill posed problem. The phenomenological basis of experimental 
FFs extraction is the well known Rosenbluth formula for $d {\sigma}/dt$ 
in the one photon exchange approximation \cite{Rosen}.  
Different experimental groups have extracted
FFs from their own data on elastic scattering with
loose information on the sources and on the nature of different errors 
estimated for differential cross sections. As a result we have no reference 
numbers to make all the data on FFs extracted by different groups mutually 
consistent.

As the common practice of usage the experimental information on FFs in 
 applications is via the ``currently best'' parametrizations 
we propose in this paper a schema to select the best FFs parametrizations by 
fitting to the world $e^{\pm} p$ elastic scattering data without using the 
derived data on FFs. 
It makes possible to have a relatively complete control of the error matrices 
of the technical and physical parameters and to estimate the errors 
induced by switching of the competitive parametrizations and errors induced 
by our method used to make world database self consistent in relative 
normalizations of different experiments.  

In this paper we analyze only $ e^\pm p$ elastic scattering data.
The article is organized as follows.

{\bf Section 1.} Here we reproduce a compilation of traditional 
formulas used in data analyses for extraction of the FFs from the data on 
$d {\sigma}/dt$.

{\bf Section 2.} presents the description of our preparatory handling of the
experimental data published so far and stored in the ReacData 
\cite{rd} database of PPDS system. Our procedure of the cross assessments of 
data and models is also outlined in this section and in the Appendices.

{\bf Section 3.} is a compilation of the parameterizations used in cross
assessments. 

{\bf Section 4.} Here we describe the preparatory fit results for each model selected 
for cross assessments. 

{\bf Section 5.} gives details of our models driven procedures to make data
from different experiments consistent with each other.

{\bf Section 6.} Here we describe the procedure of the model-dependent
filtration and present the final results of the fits to the filtered data.

{\bf Section 7.} The summary of the results obtained and the outline of the
future iterations of assessments are given in this section.

\section{General description of the $\boldmath e^{\pm}p$ elastic scattering}

The differential  cross section of the elastic $e^{\pm} p$ scattering
in the laboratory reference frame is parameterized by the well known 
Rosenbluth formula \cite{Rosen}
    
    $${ {d\sigma}\over{d\Omega} }\,=\,
 \left \{ {{d\sigma}\over{d\Omega}} \right \}_{ns}\cdot
\left [  { {\tau G_m^2 \,+\, G_E^2 } \over {1 \,+\, \tau} } 
\,+\, 2\tau G_M^2 tg^2( \theta / 2) \right ], \eqno(1) $$
\noindent
where $$\tau \,=\, {q^2\over{4m_p^2}}, \quad
\left \{ {{d\sigma}\over{d\Omega}} \right \}_{ns}
 \,=\,  \left( { \alpha\over{2E} } \right)^2 
{\cos^2(\theta / 2) \over \sin^4(\theta / 2) } \cdot
\left \{ 1 \,+\, {2E\over{m_p}}\sin^2(\theta / 2) \right \}^{-1}, $$
\noindent
$E$ is the incoming electron energy, $\theta$ is the scattering angle of the 
electron in the lab frame, and $m_p$ is the proton mass.

The Rosenbluth formula is based on the one photon exchange approximation QED 
diagram with modified proton-photon-proton
 vertex for the extended proton

 $$\Gamma_{\mu}(q^2) = \gamma_{\mu}F_1(q^2) + i\sigma_{\mu\nu}
{ {q_{\nu}}\over{2m_p} }
F_2(q^2), $$

\noindent
with Dirac ($F_1$) and Pauli ($F_2$) FFs introduced. These
FFs 
are connected with so called ``Sachs'' FFs \cite{sachs} as follows:

$$G^{p,n}_E(q^2) = F^{p,n}_1(q^2) - {{q^2}\over{4m_{p,n}^2}}F^{p,n}_2(q^2), 
\quad G^{p,n}_M(q^2) = F^{p,n}_1(q^2) + F^{p,n}_2(q^2).$$ 

\noindent
 In the non-relativistic limit they describe an electric charge
and magnetic moment distributions within the proton with the normalizations as 
follows: 
$$G^p_E(0) \,=\, 1, \qquad G^p_M(0) \,=\, {\mu}_p , $$
$$G^n_E(0) \,=\, 0, \qquad G^n_M(0) \,=\, {\mu}_n , $$ 
\noindent
where ${\mu}_{p,n}$ are the $(p,n)$ magnetic moments and
\newpage
$$ \left.{{\langle r_E^p \rangle}^2 = - 6 \cdot {\frac {dG_E^p(q^2)}{d{q^2}}}
}\right|_{q^2=0}$$
is the mean square radius of the proton charge distribution in units $[GeV^{-2}]$.
In terms of invariant variables $s$ and $t$ the formula for differential cross 
section takes the form

    $${{d\sigma}\over{d(-t)}} \,=\,
    { {4\pi{\alpha}^2}\over{t^2} }
    \left ( 1+{t\over{\tilde s}}+ {{tm_p^2}\over{{\tilde s}^2}} \right )
    { \left [ { {{G_E^2 + \tau G_M^2} \over {1 + \tau}} +
    {{G_M^2} \over {2}} {{t^2} \over {\tilde s(\tilde s + t) + tm_p^2 }} } 
     \right ] }, \eqno(2)$$

\noindent
where $\tilde s = s - m_p^2, \quad t= -q^2 = -{4E^2\sin^2(\theta/2)}\cdot
[1+{{\displaystyle 2E}\over{\displaystyle m_p}}\sin^2(\theta/2)]^{-1}$. 
Note, that $t$ varies in the range  $ -{\tilde s}^2 / s < t < 0 $.
    When $s\to\infty$, then one has $${{d\sigma}\over{d(-t)}} \to
    { {4\pi{\alpha}^2}\over{t^2} }\cdot
    { \left [ { {{G_E^2 + \tau G_M^2} \over {1 + \tau}} } \right ]},\eqno(3)$$

\noindent
for any fixed value of $t$. It can be shown that the righthand side of (3) is 
an upper envelope curve for all ${{\displaystyle d\sigma} \over
{\displaystyle d(-t)}}(s,t)$. 

As is seen from Fig. 1, this is confirmed experimentally in the wide
ranges of both variables $s$ and $t$. This is a direct evidence that
one photon approximation works well
at the available level of accuracy of the world experimental data. 

To model FFs, we follow a conventional way with introduction of the normalized
to unity at $t=0$ isospin FFs, namely,  $F^{s,v}_{1,2}$. They are 
connected with the $F^{p.n}_{1,2}$ and $G^{p,n}_{E,M}$ through the relations:

$$ F^p_1 = {1 \over 2} \left ( F^s_1 + F^v_1 \right ), \quad 
   F^p_2 = {{\tilde \mu_p + \tilde \mu_n} \over 2} F^s_2 +
           {{\tilde \mu_p - \tilde \mu_n} \over 2} F^v_2,$$

$$ F^n_1 = {1 \over 2} \left ( F^s_1 - F^v_1 \right ), \quad 
   F^n_2 = {{\tilde \mu_p + \tilde \mu_n} \over 2} F^s_2 -
           {{\tilde \mu_p - \tilde \mu_n}, \over 2} F^v_2,$$

\noindent
where $\tilde \mu_{p,n}$ are the anomalous magtnetic moments of proton and 
neutron.

\section{Data sample}

    We use the total sample of data on differential cross-sections for the 
electron-proton elastic scattering extracted from the PPDS ReacData data base 
\cite{rd} (see also Reaction Data \cite{durham} ). The complete list of 
references for the used experimental data is presented in Table~5.1.

The data are divided by the sets corresponding to different experiments or
experimental methodics. For example, if the different methodics
were used in the same work to obtain the data, we treat different data samples as 
independent because of possible difference in systematic normalization 
errors.
The data of  sets 34 and 35 are taken from the graphs.
 In all the cases, where it was possible, we separated the
systematic normalization errors from the total ones in order to 
renormalize the different data sets in a such a way that makes the whole 
compilation self-consistent. In a few cases we failed to separate 
the normalization errors
from the total errors. We classify such data sets as ``incomplete data'' and 
will use them only at the stage of the first crude adjustments of the models.
 The incomplete data sets 
are sets 2,3,18,19,23,24,41 of Table~5.1. They have only the total errors
assigned without detailed information about systematic errors. 

\begin{figure}[H]
\centerline{\epsfig{file=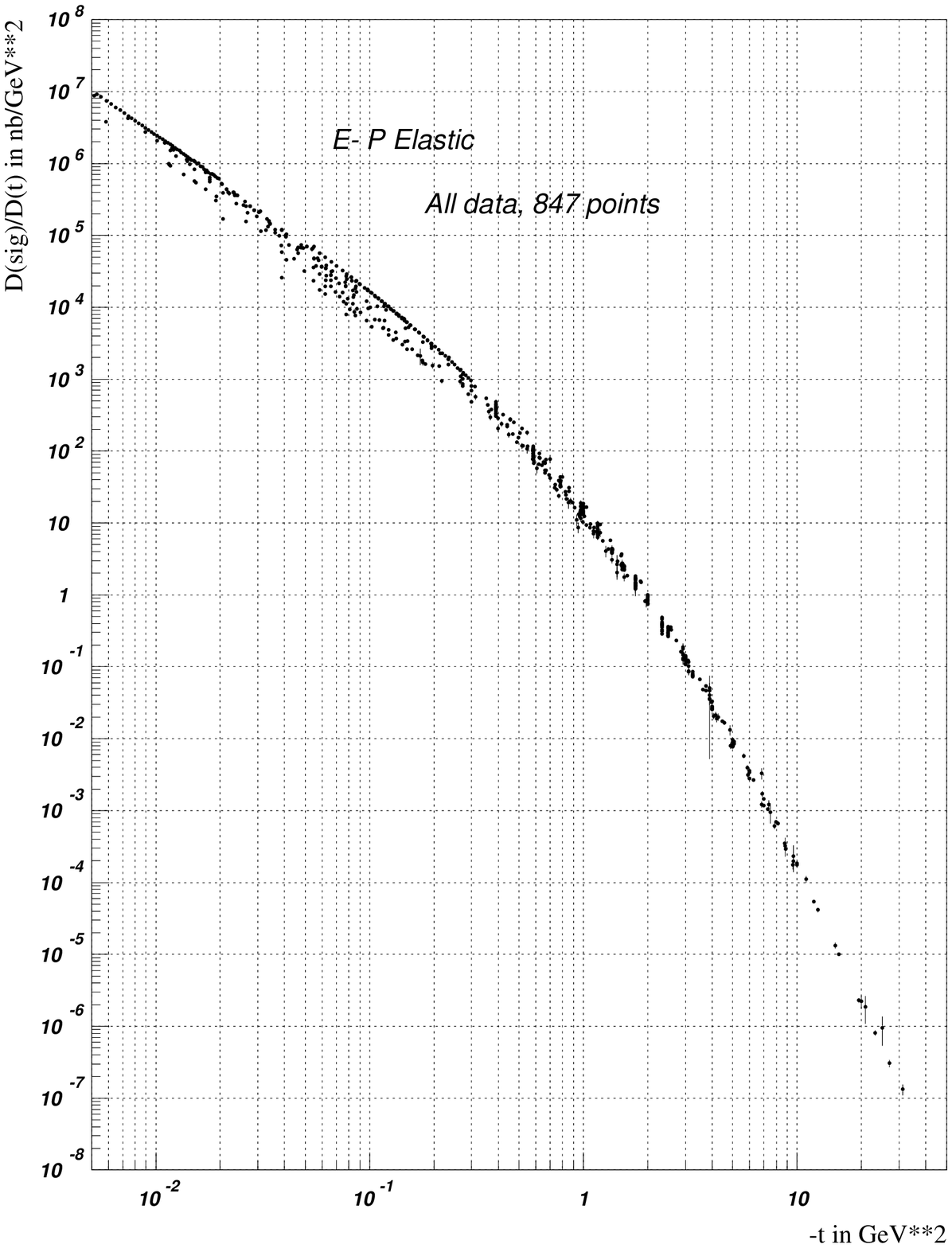,width=16cm}}
\caption{Data atlas.}
\end{figure}

\vspace*{1cm}

\begin{figure}[H]
\centerline{\epsfig{file=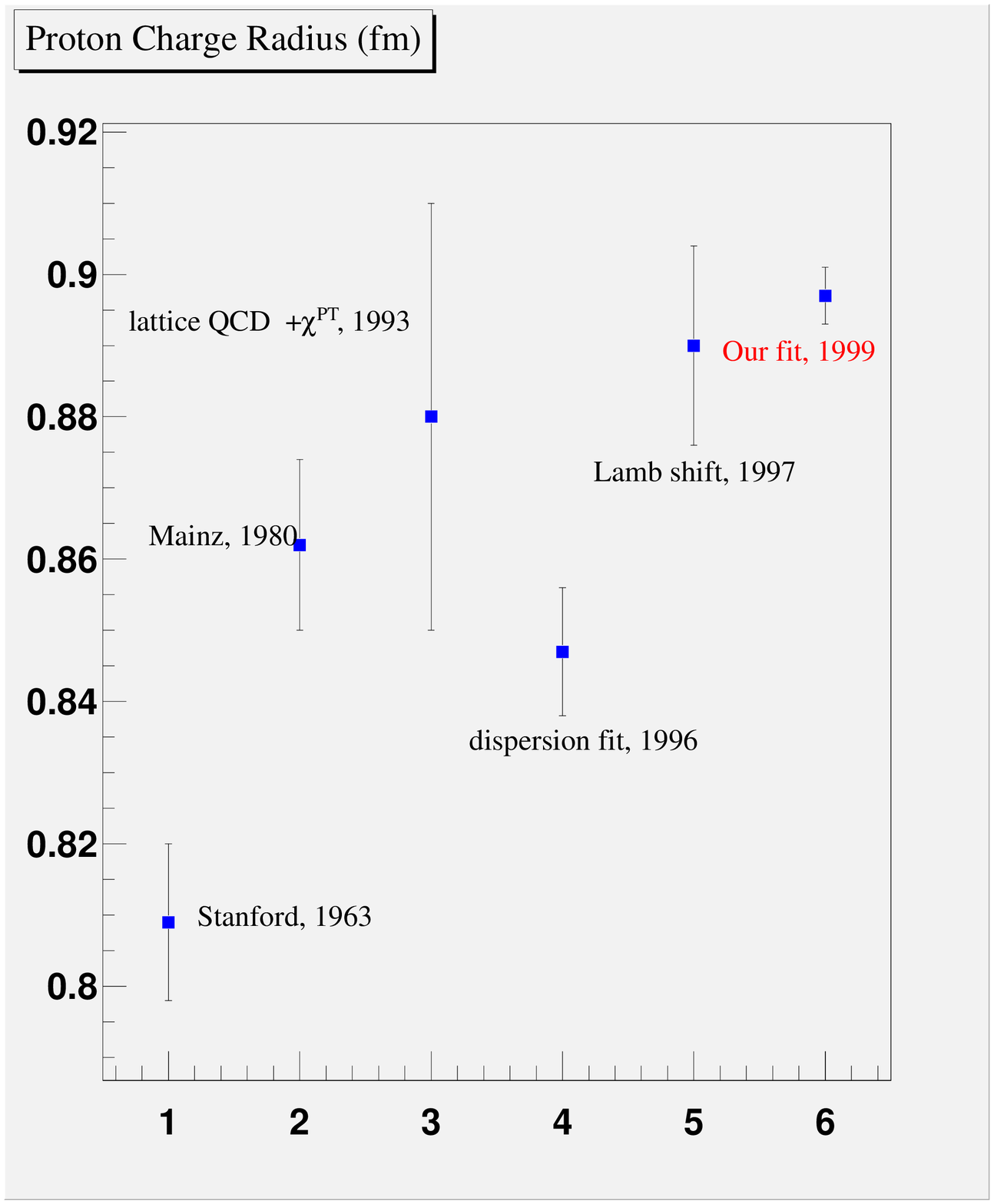,width=16cm}}
\caption{The proton charge radius as it is usually known.
(The original picture was in the article of S.S. Karshenboim,
 see hep-ph/9712347)}
\end{figure}
\newpage

 We have also excluded 5 points of set 1 ({BUMILLER 61, PR 124, 1623}, see 
Table~2.1.) that have inconsistent information about the errors, and few
points that turned out to be in sharp contradictions with the nearest experimental 
points and are out of all parametrizations by more than 5-6 standard deviations.

\small
\begin{center}
\centerline{\bf Table~2.1. Excluded points}
\vspace*{0.1cm}
\begin{tabular}{|rl|c|c|c|}
 \hline
\multicolumn{2}{|c|}{Experiment} & $q^2 [GeV^2]$ & $E [GeV]$ & $\theta [deg]$ \\ 
\hline
1 &  BUMILLER 61&    0.76 &  0.700 & 145 \\
  &             &0.45 &  0.800 & 60 \\
  &             &0.33 &  0.850 & 45 \\
  &             &0.52 &  0.875 & 60 \\
  &             &0.70 &  0.900 & 75 \\
\hline
5 &YOUNT 62 & 0.202 & 0.307 & 130 \\
\hline
6 & +YOUNT 62 & 0.202 & 0.307 & 130 \\
\hline
25& MISTRETTA 69& 0.422 & 4.529& \\ 
\hline
26& BREIDENBACH 70&  & 7. & 6 \\ 
\hline 
37& BORKOWSKY 74  & $0.76[fm^{-2}]$& 0.1498  & \\
  &                & $0.56[fm^{-2}]$& 0.1801 & \\
\hline
\end{tabular}
\end{center}
\normalsize

\noindent
The resulting data sample that we have included in the first crude analysis 
(see section 5) consists of 847 experimental points covering the  
$q^2$ range from 0.005 to 31.280 $GeV^2$ and 1961-1994 time period 
(see Fig. 1).

\section{Models}

To construct a simple but flexible enough estimator for the nucleon 
electromagnetic FFs, we explore the majority of 
previous old and recent ideas invented in the papers devoted to FFs 
modelling and data analyses \cite{hofs},\cite{clem} -- \cite{meshch97}: the 
simplest analytic structure of Pauli 
and Dirac FFs with the only established singularities that have a clear 
physical interpretation and with an asymptotic behaviour at a large 
momentum transfer as predicted by perturbative QCD \cite{brodsky}.  
Isoscalar and isovector FFs\footnote{Mathematica and Fortran codes of all
the models can be obtained by request from: polishchuk@mx.ihep.su.}~
 are modelled  as follows:
\begin{eqnarray}
F_{1,2}^{s,v}(t)& = & F_{GVMD(1,2)}^{s,v}(t) \times A_{1,2}^{s,v}(t),
\end{eqnarray}

\noindent
where $F_{GVMD(1,2)}^{s,v}(t)$ are the generalized vector dominance model 
factors, 
and $A_{1,2}^{s,v}$ are the terms which control the asymptotic behavior of FFs 
at $(-t) \to \infty$.

 All the functions used are the real analytic functions
with real magnitudes at $t < 0$, with cuts dictated by the first few particles
thresholds at $t > 0$. $F_{GVMD(1,2)}^{s,v}(t)$ have resonant poles only on 
secondary sheets of the Riemann surface of the FF and constructed as a sum of 
individual contributions corresponding to resonant poles.\\

\noindent
\fbox{\Large \boldmath $f_{pionium}(t)$}\\[-9mm] 
\hspace*{31mm} is the term~~motivated
by possible~~poles of pionium spectrum  (correspon-\-
\hspace*{31mm} ding to $ 1^{--}$ electromagnetic
bound states of the $\pi^{+} \pi^{-}$ system). There is no possibility to get 
a parametrization of the whole pionium stectrum contribution to the FFs.
Therefore, we introduce only the nearest to the point $t = 0$ pole contribution 
to the FFs
\begin{eqnarray}
f_{pionium}(t)& = &{\frac {(\sqrt{1-t/4m_{\pi}^2})\cdot
(1+\sqrt{1-(2m_{\pi}-\epsilon_{\pi})^2/4m_{\pi}^2})}
{\sqrt{1-t/4m_{\pi}^2} + \sqrt{1-(2m_{\pi}-\epsilon_{\pi})^2/4m_{\pi}^2}} .}
\end{eqnarray}

\noindent
It is the real analytic function in the whole t-plane with 
a righthand cut $[4m_{\pi}^2,\infty[$ with a pole on the second Rieman 
sheet. It corresponds to the lowest $1^{--}$ pionium ``bound'' state 
with the binding energy $-\epsilon_{\pi}$, which is unstable under annihilation.
We neglect the pionium annihilation width in form (2).
  By construction it has the following properties:   
$$f_{pionium}(0) = 1, \quad f_{pionium}(t) \to 1 \quad {\rm at} \quad 
|t| \to \infty, \quad
f_{pionium}(t)\vert_{\epsilon_{\pi}=0} \equiv  1. $$
\vspace*{0.4cm}
\noindent
\fbox{\Large \boldmath $f^{\rho}(t)$ } \\[-12mm]
\hspace*{19mm}
is the model~~of $\rho$ meson pole~~contribution to $F_{GVMD(1,2)}^{v}(t)$
~~in the whole  t-plane.\\
\hspace*{19mm}
 It is constructed from the real analytic function 
$\Phi^{\rho}(t)$, proposed by Wataghin\cite{wataghin}. It has the 
 cut $[4m_{\pi}^2, \infty[$ and two
poles in the unphysical Riemann sheets in the complex conjugated points 
\begin{eqnarray}
\Phi^{\rho}(t)& = & 
{ \frac { m_{\rho}\Gamma_{\rho} } 
{m_{\rho}^2 - t + m_{\rho}\Gamma_{\rho} \cdot 
\sqrt{ {4m_{\pi}^2-t} \over { m_{\rho}^2-4m_{\pi}^2 } } } },
\end{eqnarray}
\noindent
where $\rho$ pole positions in two complex
conjugated points are both on the unphysical sheet as required by 
real analyticity. The pole positions are  determined 
by requirement to reproduce the Breit-Wigner shape in the vicinity of $\rho$
resonance ($t=m_{\rho}^2$) at the upper edge of the cut in the small width 
approximation. 
The nonzero width $\rho$ meson  contributions to the $F_{GVMD(1,2)}^{v}(t)$
is given by 
 \begin{eqnarray}
f^{\rho}(t) & = & \frac {\Phi^{\rho}(t)}
{\Phi^{\rho)}(0)}
\end{eqnarray} 
with the following properties
$$f^{\rho}(0) = 1, \qquad f^{\rho}(t) \sim 
{m_{\rho}\Gamma_{\rho} \over t} \quad {\rm at} \quad |t| \to \infty. $$
\vspace*{-0.2cm}
\noindent
\fbox{\Large \boldmath $f_{1,2}^{\omega}(t)$ } \\[-6mm]
\hspace*{22mm} are 
the models~~of $\omega$~~meson
contributions~~to FFs in~~the whole t-plane with the \\
\hspace*{22mm} righthand cuts.  They are
the real analytic finctions with cuts $[4m_{\pi}^2, \infty[$  and 
$[9m_{\pi}^2, \infty[$   with
poles on the unphysical Riemann sheets in the complex conjugated points, and 
are constructed as follows\footnote{Here we try to test whether differrent branch
channels can be identified from the fit. So, our form for $\omega$ and $\phi$ 
contributions differs from that used in
\cite{wataghin},\cite{meshch94},\cite{meshch97}. }~~:
\begin{eqnarray}
 \Phi_{1,2}^{\omega}(t)& = & \cos{\Theta_{1,2}^{\omega}}\cdot
{ \frac { m_{\omega}\Gamma_{\omega}^{(3)} } 
{m_{\omega}^2 - t + m_{\omega}\Gamma_{\omega}/\pi  \cdot 
\ln{ {9m_{\pi}^2-t} \over { m_{\omega}^2-9m_{\pi}^2 } } } } \nonumber \\
& + & \sin{\Theta_{1,2}^{\omega}}\cdot
{ \frac {  m_{\omega}\Gamma_{\omega}^{(2)} } 
{m_{\omega}^2 - t + m_{\omega}\Gamma_{\omega} \cdot 
\sqrt{ {4m_{\pi}^2-t} \over { m_{\omega}^2-4m_{\pi}^2 } } } }, 
\end{eqnarray}
\noindent
where $\omega$ resonance pole positions are in the complex
conjugated points on the unphysical sheets corresponding to two and three body 
branch cuts. 
 The pole positions are determined 
by the requirement to reproduce the Breit-Wigner shape in the vicinity of 
resonance on the upper edge of the cut in the small width approximation. 
The nonzero width omega meson contributions to the FFs are as follows:
 \begin{eqnarray}
f_{1,2}^{\omega}(t) & = & \frac {\Phi_{1,2}^{\omega}(t)}
{\Phi_{1,2}^{\omega}(0)},
\end{eqnarray} 
with properties
$$f_{1,2}^{\omega}(0) = 1, \qquad f_{1,2}^{\omega}(t) \sim 
{ const \over t}\quad
 {\rm at} \quad |t| \to \infty .$$
\vspace*{0.1cm}

\noindent
\fbox{\Large \boldmath $f_{1,2}^{\phi}(t)$ } \\[-10mm]
\hspace*{23mm} are the
models~~of~~$\phi$~meson contributions~~to FFs in the whole~~t-plane with the\\
\hspace*{23mm}  righthand~~cuts.~~They are~~
the real~~analytic~~finctions with~~cuts~~$[4m_{K}^2, \infty[$  and\\
\hspace*{23mm}   $[9m_{\pi}^2, \infty[$ with
poles on the unphysical Riemann sheets in the complex conjugated points. 
They are constructed in a similar way
\begin{eqnarray}
 \Phi_{1,2}^{\phi}(t)& = &\cos{\Theta_{1,2}^{\phi}}\cdot
{ \frac { m_{\phi}\Gamma_{\phi}^{(3)} } 
{m_{\phi}^2 - t + m_{\phi}\Gamma_{\phi}/\pi  \cdot 
\ln{ {9m_{\pi}^2-t} \over { m_{\phi}^2-9m_{\pi}^2 } } } } \nonumber \\
& + &\sin{\Theta_{1,2}^{\phi}}\cdot
{ \frac {  m_{\phi}\Gamma_{\phi}^{(2)} } 
{m_{\phi}^2 - t + m_{\phi}\Gamma_{\phi} \cdot 
\sqrt{ {4m_{K}^2-t} \over { m_{\phi}^2-4m_{K}^2 } } } }. 
\end{eqnarray}
\noindent
The nonzero width $\phi$ meson  contributions to the FFs have the form
 \begin{eqnarray}
f_{1,2}^{\phi}(t) & = & \frac {\Phi_{1,2}^{\phi}(t)}
{\Phi_{1,2}^{\phi}(0)},
\end{eqnarray} 
with properties
$$f_{1,2}^{\phi}(0) = 1, \qquad f_{1,2}^{\phi}(t) \sim 
{ const \over t}\quad
 {\rm at} \quad |t| \to \infty .$$
\noindent
Other ingredients of our hybrid model are multiplicative 
factors $A_{1,2}^{s,v}$ which control the FFs behaviour at large 
$|t|$. We use the two forms for it.\\

\noindent
\fbox{\Large \boldmath $M_{1,2}^{s,v}(t)$}\\[-9mm] 
\hspace*{24mm} are motivated by the parametrization of Mack \cite{mack}
\begin{eqnarray}
M_i^{s,v}(t) & = & 
\left ( \frac {1} {1-t/t_{cut}^p} \right )^{\displaystyle x_1^{s,v}+i-1} 
\left ( \frac {1} {1-t/t_{cut}^{s,v} } \right )^
{\displaystyle x_2^{s,v}\ln{(t_{cut}^{s,v}-t) \over \lambda^2}}, \quad i=1,2,
\end{eqnarray} 
\noindent
where $t_{cut}^p = 4m_p^2$ is the proton-antiproton threshold, 
$t_{cut}^s = 9m_{\pi}^2,\quad t_{cut}^v = 4m_{\pi}^2$, and 
$\lambda$ is the scale parameter which is expected to be of the order of 
the $\Lambda_{QCD}$. 

$M_{1,2}^{s,v}(t)$ are the real analytic functions with singularities
only on the righthand cuts
$[t_{cut}^{s,v},\infty [$ and $[4m_p^2,\infty [$, normalized by  
 $M_{1,2}^{s,v}(0) = 1$. At large $-t$ they decrease faster 
than predicted by QCD, but they give quite good fits
at available $t$ as it will be shown further. This model is included in
the current analysis in order to test whether the QCD asymptotic is really prominent at 
the available highest $q^2$.\\ 

\noindent
\fbox{\Large \boldmath $GK_{1,2}^{s,v}(t)$} \\[-9mm] 
\hspace*{27mm} are motivated by
parametrization~~of Gari and~~Kr\"umpelman \cite{gari} with taking\\
\hspace*{27mm}  into
account the QCD asymptotic logarithms by analogy with the para\-met\-ri\-za\-tions 
used in \cite{fw},\cite{mmd}
\begin{eqnarray}
GK_i^{s,v}(t) & = & \frac {(1-t/t_{cut}^p)^{\nu+1-i} }
{ \left( 1-\eta_i^{s,v}+\eta_i^{s,v}(1-t/t_{cut}^{s,v})
\ln{\displaystyle \left( t_{cut}^{s,v}-t \over \lambda^2 \right )}/
\ln{\displaystyle t_{cut}^{s,v} \over \lambda^2}\right )^{2+\nu}}~,
\quad i=1,2,
\end{eqnarray}
\noindent
$GK_i^{s,v}(t)$ are the real analytic 
finctions with singularities only on the righthand cuts
$[t_{cut}^{s,v},\infty [$ and $[4m_p^2,\infty [$, normalized by  
 $GK_{1,2}^{s,v}(0) = 1$ at $t=0$. At large $-t$, they reproduce the 
FFs behaviour at large $-t$ as predicted by pQCD. The exponent $\nu$ 
is given \cite{brodsky} by 
$$ \nu = {4 \over {3\beta}}, \quad \beta = 11-{2 \over 3}N_f,
\quad \nu = 0.174 \quad {\rm for}\quad N_f = 5. $$
Now we are ready to assemble parametrizations for isotopis form factors.\\

\noindent
{\bf Modified Mack-67 parametrization (Mack-98)}
\begin{eqnarray}
F_{1,2}^s & = & \left( (1-C_{1,2}^{\omega})\cdot f_{pionium}(t) + 
C_{1,2}^{\omega} \cdot f_{1,2}^{\omega}(t) \right) \times M_{1,2}^s, \\
F_{1,2}^v & = & \left( (1-C_{1,2}^{\rho}) \cdot f_{pionium}(t) +
C_{1,2}^{\rho}\cdot f^{\rho}(t) \right) \times M_{1,2}^{v}
\nonumber
\end{eqnarray}
\bigskip
\noindent
with 10 adjustable parameters: $C_{1,2}^{\rho,\omega},\quad 
\Theta_{1,2}^{\omega},\quad x_{1,2}^{s,v}$.\\[-0.4cm]

\noindent
{\bf Modified Gari-Kr\"umpelman parametrization (GK-98-2V)}
\begin{eqnarray}
F_{1,2}^s & = & \left( (1-C_{1,2}^{\omega})\cdot f_{pionium}(t) + 
C_{1,2}^{\omega} \cdot f_{1,2}^{\omega}(t) \right) \times GK_{1,2}^s(t),
\nonumber \\
F_{1,2}^v & = & \left( (1-C_{1,2}^{\rho}) \cdot f_{pionium}(t) +
C_{1,2}^{\rho}\cdot f^{\rho}(t) \right) \times GK_{1,2}^{v}
\end{eqnarray}

\noindent
with 10 adjustable parameters: $C_{1,2}^{\rho,\omega},\quad 
\Theta_{1,2}^{\omega},\quad \eta_{1,2}^{s,v}$.\\

\noindent
{\bf Modified Gari-Kr\"umpelman parametrization (GK-98-3V))}
\begin{eqnarray}
F_{1,2}^s & = & \left( (1-C_{1,2}^{\omega}-C_{1,2}^{\phi})\cdot f_{pionium}(t)
 + C_{1,2}^{\omega} \cdot f_{1,2}^{\omega}(t) 
 + C_{1,2}^{\phi} \cdot f_{1,2}^{\phi}(t) \right) \times GK_{1,2}^s(t),
\nonumber \\
F_{1,2}^v & = & \left( (1-C_{1,2}^{\rho}) \cdot f_{pionium}(t) +
C_{1,2}^{\rho}\cdot f^{\rho}(t) \right) \times GK_{1,2}^{v}
\end{eqnarray}

\noindent
with 14 adjustable parameters: $C_{1,2}^{\rho,\omega,\phi},\quad 
\Theta_{1,2}^{\omega}, \quad \Theta_{1,2}^{\phi},\quad \eta_{1,2}^{s,v}$.\\

\section{Preliminary fit results}
 
 To select ``the best models'' for the FFs, we start with the crude fits 
of each above parametrization to the 
overall data sample (847 data points, see Table~5.1.). We  used a
mean least squares method with  weights defined by the ``total error'' -- the 
quadratically combined all kinds of experimental errors, ignoring  
correlations. 
The results of our best fits of analytic FFs are given in Table~4.1.
 The best fit estimations for the parameters and two kinds of their error 
estimates are presented: 1) the Minuit errors, 2) the errors obtained by 
propagation of the combined errors of the experimental data points to the 
 parameter errors using equations to find the position of minimum 
 in the parameter space. Since we will use the error propagation technique
 to get separate estimates of the parameter errors of different kinds due 
 to statistical, systematic, and  normalization errors in the input data, 
 we must be sure that we are in a local 
 minimum and parameters estimates are sufficiently close to the minimum
 position. 
  These two errors must be close to each other if the usual Minuit assumptions
 about the minimum are valid and if the distance to the true minimum is 
 sufficiently small 
 (see Appendix 2). 

\newpage
\vspace*{1.5cm}

\small
\begin{center}
\centerline{{\bf Table 4.1.} All data, weights made from all errors combined in quadratures.}
\vspace*{0.1cm}
\begin{tabular}{|c||c|c|} \hline
 Model name    & Free \& derived& Estimated parameter values and their errors\\
 \& Fit quality& parameters     &    \\
\hline
Mack-98     & $x_1^s$       & 2.702$\pm$ 0.024(0.024)\\
            & $x_1^v$       & 0.894 $\pm$ 0.032(0.031)\\
            & $x_2^s$       & 0.000 (fixed)\\
            & $x_2^v$       & 0.05232 $\pm$ 0.00047(0.00047)\\
$\chi^2/dof$& $C_1^{\rho}$  & 0.9681 $\pm$ 0.0069(0.0068)\\
986.4/836   & $C_2^{\rho}$  & 0.9127 $\pm$ 0.0049(0.0048)\\
 1.179      & $C_1^{\omega}$& 0.5701 $\pm$ 0.0071(0.0071)\\
            & $C_2^{\omega}$& $-$0.75 $\pm$ 0.17(0.17)\\
            &$\theta_1^{\omega}[\rm deg]$& 91.541$\pm$ 0.069(0.079)\\
            &$\theta_2^{\omega}[\rm deg]$& 91.41413$\pm$ 0.00013(0.00014)\\
            &$\lambda[\rm GeV]$      &  0.1024 $\pm$ 0.0039(0.0039)\\      
            & $ r_E^p[\rm fm] $ & \multicolumn{1}{c|}{
\boldmath $ 0.860 \pm 0.003(0.003) $} \\
\hline
GK-98-2V    & $\eta_1^v$       & 0.06699 $\pm$ 0.00142(0.00141)\\
            & $\eta_2^v$       & 0.0364 $\pm$ 0.0011(0.0011)\\
            & $\eta_1^s$       &0.1191 $\pm$ 0.0027(0.0027)\\
            & $\eta_2^s$       & 0.0471 $\pm$ 0.0029(0.0029)  \\
$\chi^2/dof$& $C_1^{\rho}$   & 2.380 $\pm$ 0.041(0.041)\\
973.6/836   & $C_2^{\rho}$   & 1.0843 $\pm$ 0.0093(0.0093)\\
 1.165      & $C_1^{\omega}$  & $-$1.952 $\pm$ 0.038(0.038)\\
            & $C_2^{\omega}$  & 8.01 $\pm$ 0.18(0.18)\\
            &$\theta_1^{\omega}[\rm deg]$& 91.3777 $\pm$ 0.0021(0.0021)\\
            &$\theta_2^{\omega}[\rm deg]$& 91.40405 $\pm$ 0.00053(0.00053)\\
            &$\lambda[\rm GeV]$ & 0.0133 $\pm$ 0.0050(0.0050)\\
            & $ r_E^p[\rm fm] $ & \multicolumn{1}{c|}{
\boldmath $ 0.865\pm 0.002(0.002) $} \\
\hline
GK-98-3V    & $\eta_1^v$       & 0.06806 $\pm$ 0.00099(0.00099)\\
            & $\eta_2^v$ & 0.03002 $\pm$ 0.00046(0.00047)\\
            & $\eta_1^s$       &0.10627 $\pm$ 0.00048(0.00048)\\
            & $\eta_2^s$ & 0.04876 $\pm$0.00057(0.00057)  \\
$\chi^2/dof$& $C_1^{\rho}$   & 2.312 $\pm$ 0.026(0.023)\\
972.4/833   & $C_2^{\rho}$   & 1.2383 $\pm$ 0.0070(0.0059)\\
 1.167      & $C_1^{\omega}$  & $-$2.583 $\pm$ 0.023(0.021)\\
            & $C_2^{\omega}$  & 8.99 $\pm$ 0.13(0.07)\\
            & $C_1^{\phi}$& 1.290 $\pm$ 0.022(0.022)\\
            & $C_2^{\phi}$& 1.91 $\pm$ 0.18(0.19)\\
            &$\theta_1^{\omega}[\rm deg]$& 91.3779 $\pm$ 0.0029(0.0029)\\
            &$\theta_2^{\omega}[\rm deg]$& 91.40402 $\pm$ 0.00097(0.00097)\\
            &$\theta_1^{\phi}[\rm deg]$& 157.98 $\pm$ 0.99(0.99)\\
            &$\theta_2^{\phi}[\rm deg]$& 157.95 $\pm$ 8.15(8.11)\\
            &$\lambda[\rm GeV]$ & 0.0136 $\pm$ 0.0046(0.0046)\\
            & $ r_E^p[\rm fm] $ & \multicolumn{1}{c|}{
\boldmath $ 0.861\pm 0.002(0.002) $} \\
\hline
Polynomial & $a_1$ & $-$0.241 $\pm$ 0.005(0.005) \\
$\chi^2/dof$& $a_2$ & 0.023 $\pm$ 0.009(0.009) \\
0.928      & $r_E^p[\rm fm]$ & \multicolumn{1}{c|}{\boldmath 
$ 0.849 \pm 0.008(0.008) $ }\\
\hline
\end{tabular} 
\end{center}
\normalsize

\newpage

In addition to the parametrizations under consideration we will also use the 
traditional second order polynomial 
$$1 + a_1 ({q^2 \over{4m_{\pi}^2}}) + a_2 {({q^2 \over{4m_{\pi}^2}})}^2$$ in
the $q^2$ range
from $0$ to $0.06$ $GeV^2$. Here  $4m_{\pi}^2$ = 0.078 $GeV^2$ is
the bound of convergence circle for the presentation of FF by power 
series;
$a_1$ and $a_2$ are the free parameters. This parametrization is often used to 
 estimate the proton electric radius. It is sensitive 
to the behaviour of FFs near a zero value of $q^2$ and large deviation 
of ``polynomial'' value from predictions of other models means that smooth
parametrizations 
do not reproduce a possible fine structure near $q^2=0$ and probably may 
give wrong results for proton radius. On the other hand, such a
``fine structure'' might be caused by inconsistency in normalizations of 
different data sets in this region.
 Such an inconsistency of the existing data in normalizations 
cannot be removed without using smooth parametrizations. Or we will have to 
wait until a new single experiment will precisely and densely enough measure the 
range of small $q^2$, sufficient to get reliable polynomial fits.

 It should be noted that we obtain the best quality fits of the whole 
data sample of the $d {\sigma}/dt$
by each of our three analytic models ever obtained by other analyses with other
models.
We quote in the tables the parameters errors estimated by two independent
methods as it was described above. 

 But all these fits are still not good enough to draw conclusions about the
reliability of the physical parameters estimates (in spite of that we have 
obtained the estimates of the proton electric radius in a good agreement with the 
latest published estimates \cite{mmd}).

It should be stressed also that the model with three vector mesons has no 
advantages in 
describing the data in the space-like region of these naive fits.  Pionium,
$\rho$, and $\omega$ contributions are sufficient to get a relatively good data
description.

Let us try to get more reliable fits by excluding correlated normalization 
errors from the weights, because this may cause (as we will show further) a 
significant biases in parameter estimates. In the 
next section we use the correlated 
normalization uncertainties to make data sets relatively consistent
with each other on the basis of the selected models.  

\section{Data normalizing}
For further analyses we have excluded  
the so called ``incomplete data sets'': 2, 3, 18, 19, 23, 24, 41 
from the total sample of the data (see Table~5.1).
They do not contain any information about systematics, and cannot be used 
in the procedure of renormalizations. To uncover a possible effect of this
formal filtration we repeate the previous fit on the reduced sample of 
experimental data with the results presented in Table~5.2. As is seen we
obtain quite good fit qualities for all the models and parameters shifted
only slightly. As for the proton radii, they are not practically shifted.

Having obtained a good enough fits by all the parameterizations, we 
remain on the firm ground to discuss the values of the proton radius. As is seen from
 Table~5.2, the model dependent radii are close to each other with
overlapping
error bands, but the radius obtained from a polynomial fit is smaller and its interval
estimator, $$ r^p_E \in [0.841,0.857] [\rm fm], $$
 does not overlap with a joined interval estimator of model radii, namely
$$ r^p_E \in [0.858,0.867] [\rm fm]. $$

\newpage

\small
\begin{center}
    {\bf Table 5.1. Statistics of data sets, renormalization factors,  and
references.}\\
\footnotesize
\vspace*{0.1cm}
\begin{tabular}{|r||c|c|c|c|r|l|} \hline
  Set & Set   & After     &  Normalization & Points in&    &          \\ 
  No  &number of& filtration &  factors       &90\% CL & 
REACDATA ShortCode&Reference \\
      & points  &                   & & sample   &         &          \\
\hline  
     1& 53  & 53  & $ 0.985 \pm  0.002$& 45  &BUMILLER 61     &  PR 124, 1623 \\
     2&  9  &  0  &         --         &0 & OLSON 61       &  PRL 6, 286 \\
     3& 20  &  0  &         --         &0 & LITTAUER 61    &  PRL 7, 141 \\
     4&  2  &  2  & $ 0.999 \pm  0.003$&0 & LEHMANN 62     &  PR 126, 1183 \\
     5&  3  &  3  & $ 1.009 \pm  0.004$&3 & YOUNT 62       &  PR 128, 1842 \\
     6&  3 &  3   & $ 1.000 \pm  0.000$&3 &+ YOUNT 62      &  PR 128, 1842 \\
     7&  8 &  8   & $ 1.017 \pm  0.004$&7 & DRICKEY 62     &  PRL 9, 521 \\
     8&  6 &  6   & $ 1.006 \pm  0.001$&4 & DUDELZAK 63    &  NC 28, 18 \\
     9& 21 & 21   & $ 1.011 \pm  0.011$&17 & BERKELMAN 63   &  PR 130, 2061 \\
    10&  7 &  7   & $ 0.988 \pm  0.013$&5 & DUNNING 63     &  PRL 10, 500 \\
    11&  6 &  6   & $ 1.030 \pm  0.012$&5 & CHEN 65        &  PR 141, 1267 \\
    12& 12 & 12   & $ 1.052 \pm  0.009$&8 &+ CHEN 65       &  PR 141, 1267 \\
    13& 10 & 10   & $ 1.000 \pm  0.000$&8 & FREREJACQUE 65 &  PR 141, 1308 \\
    14&  3 &  3   & $ 1.008 \pm  0.006$&3 & GROSSETETE 65  &  PR 141, 1435 \\
    15& 93 & 93   & $ 1.000 \pm  0.000$&88 & JANSSENS 66    &  PR 142, 922 \\
    16&  3 &  3   & $ 1.022 \pm  0.005$&0 & BENAKSAS 66    &  PR 148, 1327 \\
    17& 11 & 11   & $ 1.000 \pm  0.000$&10 & BARTEL 66      &  PRL 17, 608 \\
    18& 10 &  0   &         --         &0 & ALBRECHT 66    &  PRL 17, 1192 \\
    19&  7 &  0   &         --         &0 &+ ALBRECHT 66   &  PRL 17, 1192 \\
    20& 24 & 24   & $ 0.986 \pm  0.004$&20 & BEHREND 67     &  NC A48, 140 \\
    21&  4 &  4   & $ 0.912 \pm  0.013$&1 &+ BEHREND 67    &  NC A48, 140 \\
    22&  1 &  1   & $ 1.009 \pm  0.015$&1 & BARTEL 67      &  PL 25B, 236 \\
    23&  4 &  0   &         --         &0 & BARTEL 67B     &  PL 25B, 242 \\
    24&  3 &  0   &         --         &0 & ALBRECHT 67    &  PRL 18, 1014 \\
    25&  4 &  4   & $ 0.997 \pm  0.003 $&3 & MISTRETTA 69   &  PR 184, 1487 \\
    26&  8 &  8   & $ 1.003 \pm  0.009 $&8 & BREIDENBACH 70 &  MIT-2098-635 \\
    27& 22 & 22   & $ 1.004 \pm  0.003$&18 & LITT 70        &  PL 31B, 40 \\
    28& 15 & 15   & $ 1.002 \pm  0.003$&11 & GOITEIN 70     &  PR D1, 2449 \\
    29& 54 & 54   & $ 0.996 \pm  0.002$&52 & BERGER 71      &  PL 35B, 87 \\
    30&  9 &  9   & $ 1.005 \pm  0.003$&9 & PRICE 71B      &  PR D4, 45 \\
    31&  4 &  4   & $ 0.985 \pm  0.012$&4 & GANICHOT 72    &  NP A178, 545 \\
    32&  5 &  5   & $ 1.003 \pm  0.004$&5 & BARTEL 73      &  NP B58, 429 \\
    33& 16 & 16   & $ 0.997 \pm  0.003$&15 & + BARTEL 73    &  NP B58, 429 \\
    34& 37 & 37   & $ 0.992 \pm  0.001$&35 & BOTTERILL 73C  &  PL 46B, 125 \\
    35& 30 & 30   & $ 0.988 \pm  0.001$&22 &+ BOTTERILL 73C &  PL 46B, 125 \\
    36& 11 & 11   & $ 0.997 \pm  0.006$&10 & KIRK 73        &  PR D8, 63 \\
    37& 44 & 44   & $ 0.994 \pm  0.001$&37 &  BORKOWSKY 74  &  NP A222, 269 \\
    38& 11 &  11   & $ 1.013 \pm  0.002$&11 &  MURPHY 74     &  PR C9, 2125 \\
    39& 77 & 77   & $ 0.996 \pm  0.001$&69 &  BORKOWSKY 75  &  NP B93, 461 \\
    40&  7 &  7   & $ 0.993 \pm  0.004$&7 &  STEIN 75      &  PR D12, 1884 \\
    41&  3 &  0   &         --         &0 &  MARTIN 76C    &  PRL 38, 1320 \\
    42&  8 &  8   & $ 1.022 \pm  0.008$&8 &  MESTAYER 78   &  SLAC-214 \\
    43& 49 & 49   & $ 1.002 \pm  0.000$&40 &  AKIMOV 78B    &  YF 29, 922 \\
    44&  5 &  5   & $ 1.002 \pm  0.001$&5 &  SIMON 80      &  NP A333, 381 \\
    45& 22 & 22   & $ 0.998 \pm  0.002$&21 &  WALKER 89     &  PL 224B, 353 \\
    46& 11 &  11   & $ 0.995 \pm  0.005$&11 &  BOSTED 89     &  PR C42, 38 \\
    47&  5 &  5   & $ 0.969 \pm  0.011$&5 &  ROCK 91       &  PR D46, 24 \\
    48& 13 & 13   & $ 0.994 \pm  0.004$&12 &  SILL 93       &  PR D48, 29 \\
    49& 22 & 22   & $ 0.998 \pm  0.002$&22 &  WALKER 94     &  PR D49, 5671 \\
    50& 32 & 32   & $ 1.000 \pm  0.002$&28 &  ANDIVAHIS 94  &  PR D50, 5491 \\
\hline
   All&{\bf 847} &{\bf 791} &            &{\bf 696} & & \\ 
\hline
\end{tabular}

\end{center}


\newpage

\vspace*{1cm}

\small
\centerline{{\bf Table 5.2.} Fits to sets of complete data only, weights made from all errors
combined in quadratures.}
\begin{center}
\begin{tabular}{|c||c|c|} \hline
 Model name    & Free \& derived& Estimated parameter values and their errors\\
 \& Fit quality& parameters     &  \\   
\hline
Mack-98     & $x_1^s$       & 2.587 $\pm$ 0.091(0.079)\\
            & $x_1^v$       & 0.978 $\pm$ 0.072(0.061)\\
            & $x_2^s$       & 0.0035 $\pm$ 0.0029(0.0027)\\
            & $x_2^v$       & 0.0505 $\pm$ 0.0017(0.0015)\\
            & $C_1^{\rho}$  & 0.9665 $\pm$ 0.0072(0.0071)\\
$\chi^2/dof$& $C_2^{\rho}$  & 0.9063 $\pm$ 0.0051(0.0050)\\
757.4/780   & $C_1^{\omega}$& 0.5767 $\pm$ 0.0073(0.0072)\\
 0.971      & $C_2^{\omega}$& $-$0.97 $\pm$ 0.17(0.17)\\
            &$\theta_1^{\omega}[\rm deg]$& 91.523 $\pm$ 0.045(0.044)\\
            &$\theta_2^{\omega}[\rm deg]$& 91.414162 $\pm$ 0.000085(0.000084)\\
            &$\lambda[\rm GeV]$      &  0.1010 $\pm$ 0.0039(0.0039)\\      
            & $ r_E^p[\rm fm] $ & {
\boldmath $ 0.861\pm 0.003(0.003) $} \\
\hline
GK-98-2V    & $\eta_1^v$       & 0.0624 $\pm$ 0.0015(0.0015)\\
            & $\eta_2^v$ & 0.02762 $\pm$ 0.00090(0.00090)\\
            & $\eta_1^s$       &0.1094 $\pm$ 0.0027(0.0027)\\
            & $\eta_2^s$ & 0.0427 $\pm$0.0023(0.0023)  \\
$\chi^2/dof$& $C_1^{\rho}$   & 2.317 $\pm$ 0.038(0.039)\\
750.4/780   & $C_2^{\rho}$   & 1.1894 $\pm$ 0.0090(0.0090)\\
 0.962      & $C_1^{\omega}$  & $-$1.723 $\pm$ 0.035(0.035)\\
            & $C_2^{\omega}$  & 9.01 $\pm$ 0.19(0.19)\\
            &$\theta_1^{\omega}[\rm deg]$& 91.3779 $\pm$ 0.0023(0.0023)\\
            &$\theta_2^{\omega}[\rm deg]$& 91.40402 $\pm$ 0.00047(0.00047)\\
            &$\lambda[\rm GeV] $ &0.0194 $\pm$ 0.0064(0.0064)\\
            & $ r_E^p[\rm fm] $ & \multicolumn{1}{c|}{
\boldmath $ 0.865\pm 0.002(0.002) $} \\
\hline
GK-98-3V    & $\eta_1^v$       & 0.0505 $\pm$ 0.0014(0.0014)\\
            & $\eta_2^v$ & 0.02281 $\pm$ 0.00072(0.00073)\\
            & $\eta_1^s$       &0.0989 $\pm$ 0.0020(0.0020)\\
            & $\eta_2^s$ & 0.0437 $\pm$0.0017(0.0017)  \\
$\chi^2/dof$& $C_1^{\rho}$   & 3.3392 $\pm$ 0.0092(0.0092)\\
747.9/776   & $C_2^{\rho}$   & 1.3402 $\pm$ 0.0028(0.0028)\\
 0.963      & $C_1^{\omega}$  & $-$3.9995 $\pm$ 0.0092(0.0106)\\
            & $C_2^{\omega}$  & 9.000 $\pm$ 0.095(0.109)\\
            & $C_1^{\phi}$& 1.9176 $\pm$ 0.0061(0.0038)\\
            & $C_2^{\phi}$& 3.999 $\pm$ 0.056(0.034)\\
            &$\theta_1^{\omega}[\rm deg]$& 91.37789 $\pm$ 0.00099 (0.00099)\\
            &$\theta_2^{\omega}[\rm deg]$& 91.40399 $\pm$ 0.00051 (0.00051)\\
            &$\theta_1^{\phi}[\rm deg]$& 157.99  $\pm$ 0.68(0.68)\\
            &$\theta_2^{\phi}[\rm deg]$& 158.0$\pm$ 3.7 (3.7)\\
            &$\lambda[\rm GeV] $ &0.0203 $\pm$ 0.0059(0.0060)\\
            & $ r_E^p[\rm fm] $ & \multicolumn{1}{c|}{
\boldmath $ 0.862\pm 0.002(0.002) $} \\
\hline
Polynomial & $a_1$ & $-$0.2407 $\pm$ 0.0047(0.0047) \\
$\chi^2/dof$& $a_2$ & 0.0229 $\pm$ 0.0088(0.0089) \\
0.928      & $r_E^p[\rm fm]$ & \multicolumn{1}{c|}{ 
\boldmath $ 0.849 \pm 0.008(0.008) $ }\\
\hline
\end{tabular}
\end{center}

\newpage

\vspace*{1cm}

\small
\centerline{{\bf Table 5.3.} Fits to sets of complete data only, weights constracted without
correlated normalization errors.}
\begin{center}
\begin{tabular}{|c||c|c|} \hline
 Model name    & Free \& derived& Estimated parameter values and their errors\\
 \& Fit quality & parameters       &    \\   
\hline
Mack-98     & $x_1^s$       & 2.03 $\pm$ 0.17(0.15)\\
            & $x_1^v$       & 1.89 $\pm$ 0.02(0.02)\\
            & $x_2^s$       & 0.0755 $\pm$ 0.0055(0.0049)\\
            & $x_2^v$       & 0.0225 $\pm$ 0.0015(0.0013)\\
$\chi^2/dof$& $C_1^{\rho}$  & $-$0.2123 $\pm$ 0.0087(0.0086)\\
1474.1/780  & $C_2^{\rho}$  & 0.6689 $\pm$ 0.0058(0.0057)\\
 1.889      & $C_1^{\omega}$& $-$1.699 $\pm$ 0.016(0.016)\\
            & $C_2^{\omega}$& $-$9.37 $\pm$ 0.25(0.25)\\
            &$\theta_1^{\omega}[\rm deg]$& 91.543 $\pm$ 0.017(0.017)\\
            &$\theta_2^{\omega}[\rm deg]$& 91.414124 $\pm$ 0.000017(0.000017)\\
            &$\lambda[\rm GeV]$      &  0.00105 $\pm$ 0.00040(0.00036)\\      
            & $ r_E^p[\rm fm] $ & \multicolumn{1}{c|}{
\boldmath $ 0.909\pm 0.001(0.001) $} \\
\hline
GK-98-2V    & $\eta_1^v$      &0.00589 $\pm$0.00012(0.00007)\\
            & $\eta_2^v$      &0.004263 $\pm$ 0.000023(0.000041)\\ 
            & $\eta_1^s$      &0.03401 $\pm$ 0.00064(0.00035)\\
            & $\eta_2^s$      &0.01211 $\pm$ 0.00043(0.00040)\\
$\chi^2/dof$& $C_1^{\rho}$     &$-$10.47 $\pm$ 0.19(0.10)\\
1447.7/780  & $C_2^{\rho}$     &0.478 $\pm$ 0.021(0.019)\\
 1.856      & $C_1^{\omega}$  &$-$9.3037 $\pm$ 0.0041(0.0060)\\
            & $C_2^{\omega}$  &1.567 $\pm$ 0.047(0.042)\\
            & $\theta_1^{\omega}[\rm deg]$& 91.504 $\pm$ 0.011(0.010)\\
            & $\theta_2^{\omega}[\rm deg]$& 91.415123 $\pm$ 0.000002(0.000003)\\
            & $\lambda[\rm GeV]$ & 0.219000 $\pm$ 0.000012(0.000012)\\
            &$r_E^p[\rm fm]  $& \multicolumn{1}{c|}{
\boldmath $  0.9115 \pm 0.0018(0.0014) $} \\
\hline
GK-98-3V    & $\eta_1^v$       &0.005887 $\pm$ 0.000028(0.000028)\\
            & $\eta_2^v$ & 0.004263 $\pm$ 0.000062(0.000063)\\
            & $\eta_1^s$       &0.03401 $\pm$ 0.00014(0.00014)\\
            & $\eta_2^s$ & 0.012101 $\pm$ 0.00044(0.00045)  \\
$\chi^2/dof$& $C_1^{\rho}$   & 10.47 $\pm$ 0.21(0.12)\\
1447.7/776  & $C_2^{\rho}$   & 0.476 $\pm$ 0.034(0.032)\\
 1.866      & $C_1^{\omega}$  & $-$9.31 $\pm$ 0.27(0.16)\\
            & $C_2^{\omega}$  & 1.571 $\pm$ 0.074(0.072)\\
            & $C_1^{\phi}$& 0.0015 $\pm$ 0.0823(0.0525)\\
            & $C_2^{\phi}$& $-$0.02 $\pm$ 0.27(0.26)\\
            &$\theta_1^{\omega}[\rm deg]$& 91.50415 $\pm$ 0.00067 (0.00067)\\
            &$\theta_2^{\omega}[\rm deg]$& 91.415123 $\pm$ 0.000001 (0.000001)\\
            &$\theta_1^{\phi}[\rm deg]$& 157.9  $\pm$ 3.7(3.7)\\
            &$\theta_2^{\phi}[\rm deg]$& 158.0$\pm$ 17.5 (17.9)\\
            &$\lambda[\rm GeV] $ &0.2190 $\pm$ 0.0059(0.0060)\\
            & $ r_E^p[\rm fm] $ & {
\boldmath $ 0.911\pm 0.001(0.001) $} \\
\hline
Polynomial  & $a_1$ & $-$0.2542 $\pm$ 0.0030(0.0030) \\
$\chi^2/dof$& $a_2$ & 0.0372 $\pm$ 0.0050(0.0050) \\
2.187       & $r_E^p[\rm fm]$ & { 
\boldmath $ 0.873 \pm 0.005(0.005) $ }\\
\hline
\end{tabular}
\end{center}
\newpage

\small
\centerline{{\bf Table 5.4.} Fits with set of normalizations determined for each model.}
\begin{center}
\vspace*{0.1cm}
\footnotesize
\begin{tabular}{|c||c|c|} \hline
 Model name   & Free and  & \multicolumn{1}{c|}{ }  \\
 and          & derived   & \multicolumn{1}{c|}{
Estimated parameter values and their errors }       \\
 fit quality  & parameters  & \multicolumn{1}{c|}{               } \\
\hline
Mack-98     & $x_1^s$       & 1.94 $\pm$ 0.17(0.15) $\pm$ 0.06(norm)\\
            & $x_1^v$       & 1.85 $\pm$ 0.020(0.019) $\pm$ 0.008(norm)\\
            & $x_2^s$       & 0.0863 $\pm$ 0.0057(0.0052) $\pm$ 0.0023(norm)\\
            & $x_2^v$       & 0.0256 $\pm$ 0.0014(0.0013) $\pm$ 0.0006(norm)\\
$\chi^2/dof$& $C_1^{\rho}$  & $-$0.2035 $\pm$ 0.0087(0.0086) 
              $\pm$ 0.0026(norm)\\
1195.6/780  & $C_2^{\rho}$  & 0.6660 $\pm$ 0.0058(0.0058) $\pm$ 0.0017(norm)\\
 1.533      & $C_1^{\omega}$& $-$1.744 $\pm$0.015(0.015) $\pm$ 0.008(norm)\\
            & $C_2^{\omega}$& $-$8.74 $\pm$ 0.25(0.25) $\pm$ 0.09(norm)\\
            &$\theta_1^{\omega}[\rm deg]$& 91.519 $\pm$ 0.011(0.011) $\pm$
            0.009(norm)\\
            &$\theta_2^{\omega}[\rm deg]$& 91.414116 $\pm$0.000019(0.000019)
            $\pm$ 0.000009(norm)\\
            &$\lambda[\rm GeV]$      &  0.00229 $\pm$ 0.00067(0.00061) $\pm$ 
            0.00028(norm)\\      
            & $ r_E^p[\rm fm] $ & {
\boldmath $ 0.8950\pm 0.0014(0.0013) \pm 0.0009$ \bf (norm) } \\
\hline
Polynomial     & $a_1$ & $-$0.2582 $\pm$ 0.0030(0.0030) $\pm$ 0.0560(norm) \\
$\chi^2/dof$& $a_2$ & 0.0505$\pm$ 0.0050(0.0050) $\pm$ 0.0695(norm) \\
1.603      &$r_E^p[\rm fm]$ & { 
\boldmath $ 0.8794 \pm 0.0051(0.0051) \pm 0.0954$ \bf (norm) }\\
\hline
GK-98-2V  &$\eta_1^v$   &0.005646 $\pm$ 0.000028(0.000029) 
           $\pm$ 0.000280(norm)\\
          &$\eta_2^v$  &0.004363 $\pm$ 0.000068(0.000072)
           $\pm$ 0.000185(norm)\\
          &$\eta_1^s$ & 0.03277$\pm$ 0.00013(0.00014) $\pm$ 0.00150(norm)\\
            &$\eta_2^s$    &0.01262 $\pm$ 0.00050(0.00054) $\pm$ 0.00155(norm)\\
$\chi^2/dof$&$C_1^{\rho}$   &9.457 $\pm$ 0.074(0.080) $\pm$ 0.578(norm)\\
 1177.9/780 &$C_2^{\rho}$   &0.4516 $\pm$ 0.0075(0.0080) $\pm$ 0.0284(norm)\\
 1.510      &$C_1^{\omega}$ &$-$8.213 $\pm$ 0.079(0.084) $\pm$ 0.627(norm)\\
            &$C_2^{\omega}$ & 1.617 $\pm$ 0.019(0.020) $\pm$ 0.074(norm)\\
            &$\theta_1^{\omega}[\rm deg]$& 91.50423 $\pm$ 0.00079(0.00079)
             $\pm$ 0.00862(norm) \\
            &$\theta_2^{\omega}[\rm deg]$& 91.415123  $\pm$ 0.000001(0.000001)
             $\pm$ 0.000003(norm)\\
            &$\lambda[\rm GeV]$      &  0.219637 $\pm$ 0.000033(0.000033) $\pm$ 
             0.000500(norm)\\      
            &$r_E^p[\rm fm]  $& {
\boldmath $  0.8986 \pm 0.0012(0.0016) \pm 0.0041$ \bf (norm) } \\
\hline
Polynomial     & $a_1$ & $-$0.2598 $\pm$ 0.0030(0.0030) $\pm$ 0.0560(norm) \\
$\chi^2/dof$& $a_2$ & 0.0531$\pm$ 0.0050(0.0050) $\pm$ 0.0695(norm) \\
1.605      &$r_E^p[\rm fm]$ & { 
\boldmath $ 0.8822 \pm 0.0051(0.0051) \pm 0.0951$ \bf (norm) }\\
\hline
GK-98-3V    & $\eta_1^v$       &0.005343 $\pm$ 0.000026(0.000026)
            $\pm$ 0.000021(norm)\\
            & $\eta_2^v$ & 0.003972 $\pm$ 0.000044(0.000044)
            $\pm$ 0.000020(norm)\\
            & $\eta_1^s$       &0.03281 $\pm$ 0.00013(0.00013) $\pm$ 
            0.00010(norm)\\
            & $\eta_2^s$ & 0.01221 $\pm$ 0.00030(0.00030) $\pm$
            0.00014(norm)\\
$\chi^2/dof$& $C_1^{\rho}$   & 9.285 $\pm$ 0.070(0.069) $\pm$ 0.045(norm)\\
1176.1/776  & $C_2^{\rho}$   & 0.3355 $\pm$ 0.0070(0.0069) $\pm$ 0.0033(norm)\\
 1.516      & $C_1^{\omega}$  & $-$9.573 $\pm$0.074(0.073) $\pm$ 0.046(norm)\\
            & $C_2^{\omega}$  & 1.778 $\pm$ 0.018(0.018) $\pm$0.009(norm)\\
            & $C_1^{\phi}$& 1.3524 $\pm$ 0.0024(0.0024) $\pm$ 0.0019(norm)\\
            & $C_2^{\phi}$& $-$1.255 $\pm$ 0.041(0.041) $\pm$ 0.015(norm)\\
            &$\theta_1^{\omega}[\rm deg]$& 91.50414 $\pm$ 0.00066 (0.00066)
            $\pm$ 0.00057(norm)\\
            &$\theta_2^{\omega}[\rm deg]$& 91.415123 $\pm$ 0.000001 (0.000001)
            $\pm$ 0.000001(norm)\\
            &$\theta_1^{\phi}[\rm deg]$& 162.4313  $\pm$ 0.0011(0.0011)
            $\pm$ 0.0009(norm)\\
            &$\theta_2^{\phi}[\rm deg]$& 162.370$\pm$ 0.010(0.010)
            $\pm$ 0.007(norm)\\
            &$\lambda[\rm GeV] $ &0.224153 $\pm$ 0.000056(0.000056) 
            $\pm$0.000023(norm) \\
            & $ r_E^p[\rm fm] $ & {
\boldmath $ 0.9029\pm 0.0009(0.0009) \pm 0.0007$ \bf(norm) } \\
\hline
Polynomial     & $a_1$ & $-$0.2607 $\pm$ 0.0030(0.0030) $\pm$ 0.0560(norm) \\
$\chi^2/dof$& $a_2$ & 0.0525$\pm$ 0.0050(0.0050) $\pm$ 0.0695(norm) \\
1.650      &$r_E^p[\rm fm]$ & \multicolumn{1}{c|}{ 
\boldmath $ 0.8838 \pm 0.0051(0.0051) \pm 0.0949$ \bf (norm) }\\
\hline
\end{tabular}
\end{center}

\newpage

\vspace*{1cm}

\small
\centerline{{\bf Table 5.5.} Fits with common set of averaged normalizations.}
\begin{center}
\vspace*{0.1cm}
\begin{tabular}{|c||c|c|} \hline
 Model name     & Free \& derived& Estimated parameter values and their errors\\
 \& Fit quality & parameters     &    \\   
\hline
Mack-98     & $x_1^s$       & 2.03 $\pm$ 0.14(0.10) $\pm$ 0.03(norm)\\
            & $x_1^v$       & 1.866 $\pm$ 0.017(0.014) $\pm$ 0.005(norm)\\
            & $x_2^s$       & 0.0819 $\pm$ 0.0046(0.0035) $\pm$0.0016(norm)\\
            & $x_2^v$       & 0.0243 $\pm$ 0.0011(0.0008) $\pm$ 0.0003(norm)\\
$\chi^2/dof$& $C_1^{\rho}$  & $-$0.2069 $\pm$ 0.0087(0.0086) 
              $\pm$ 0.0026(norm)\\
1199.8/780  & $C_2^{\rho}$  & 0.6682 $\pm$ 0.0058(0.0058) $\pm$ 0.0017(norm)\\
 1.538      & $C_1^{\omega}$& $-$1.737 $\pm$0.015(0.015) $\pm$ 0.008(norm)\\
            & $C_2^{\omega}$& $-$8.88 $\pm$ 0.25(0.25) $\pm$ 0.09(norm)\\
            & $\theta_1^{\omega}[\rm deg]$& 91.520 $\pm$ 0.011(0.011) $\pm$
              0.009(norm)\\
            & $\theta_2^{\omega}[\rm deg]$& 91.414116 $\pm$0.000019(0.000019)
              $\pm$ 0.000009(norm)\\
            & $\lambda[\rm GeV]$      & 0.00174 $\pm$ 0.00046(0.00035) $\pm$ 
              0.00017(norm)\\      
            & $ r_E^p[\rm fm] $ & {
\boldmath $ 0.8975\pm 0.0012(0.0010) \pm 0.0008$ (norm)}  \\
\hline
GK-98-2V & $\eta_1^v$ &0.005690 $\pm$ 0.000028(0.000029) $\pm$ 0.000023(norm)\\
            & $\eta_2^v$  &0.004326 $\pm$ 0.000064(0.000066)
              $\pm$ 0.000028(norm)\\
            & $\eta_1^s$ & 0.03299$\pm$ 0.00013(0.00013) $\pm$ 0.00011(norm)\\
            & $\eta_2^s$  &0.01239 $\pm$ 0.00047(0.00049) $\pm$ 0.00021(norm)\\
$\chi^2/dof$& $C_1^{\rho}$&9.53 $\pm$ 0.17(0.15) $\pm$ 0.08(norm)\\
 1179.5/780 &$C_2^{\rho}$ &0.4575 $\pm$ 0.0078(0.0078) $\pm$ 0.0034(norm)\\
 1.512      &$C_1^{\omega}$ &$-$8.29 $\pm$ 0.18(0.16) $\pm$ 0.09(norm)\\
            &$C_2^{\omega}$ & 1.601 $\pm$ 0.024(0.023) $\pm$ 0.011(norm)\\
            &$\theta_1^{\omega}[\rm deg]$& 91.50425 $\pm$ 0.00075(0.00075)
             $\pm$ 0.00065(norm) \\
            &$\theta_2^{\omega}[\rm deg]$& 91.415123  $\pm$ 0.000001(0.000001)
             $\pm$ 0.000001(norm)\\
            &$\lambda[\rm GeV]$      & 0.21963 $\pm$ 0.00019(0.00017) $\pm$ 
             0.00008(norm)\\      
            &$r_E^p[\rm fm]  $ & {
\boldmath $  0.8999 \pm 0.0014(0.0013) \pm 0.0006$ (norm) } \\
\hline
GK-98-3V    & $\eta_1^v$       &0.005314 $\pm$ 0.000025(0.000025)
              $\pm$ 0.000020(norm)\\
            & $\eta_2^v$ & 0.003992 $\pm$ 0.000046(0.000046)
              $\pm$ 0.000021(norm)\\
            & $\eta_1^s$       &0.03267 $\pm$ 0.00012(0.00012) $\pm$ 
              0.00010(norm)\\
            & $\eta_2^s$ & 0.01236 $\pm$ 0.00032(0.00032) $\pm$
              0.00015(norm)\\
$\chi^2/dof$& $C_1^{\rho}$   & 9.27 $\pm$ 0.13(0.25) $\pm$ 0.16(norm)\\
1173.8/776  & $C_2^{\rho}$   & 0.334 $\pm$ 0.012(0.023) $\pm$ 0.013(norm)\\
 1.513      & $C_1^{\omega}$  & $-$9.56 $\pm$0.14(0.27) $\pm$ 0.17(norm)\\
            & $C_2^{\omega}$  & 1.783 $\pm$ 0.029(0.051) $\pm$0.031(norm)\\
            & $C_1^{\phi}$& 1.3524 $\pm$ 0.0024(0.0024) $\pm$ 0.0018(norm)\\
            & $C_2^{\phi}$& $-$1.261 $\pm$ 0.042(0.042) $\pm$ 0.015(norm)\\
            & $\theta_1^{\omega}[\rm deg]$& 91.50412 $\pm$ 0.00066 (0.00066)
              $\pm$ 0.00057(norm)\\
            & $\theta_2^{\omega}[\rm deg]$& 91.415123 $\pm$ 0.000001 (0.000001)
              $\pm$ 0.000001(norm)\\
            & $\theta_1^{\phi}[\rm deg]$& 162.4313  $\pm$ 0.0011(0.0011)
              $\pm$ 0.0009(norm)\\
            & $\theta_2^{\phi}[\rm deg]$& 162.371$\pm$ 0.010(0.010)
              $\pm$ 0.007(norm)\\
            & $\lambda[\rm GeV] $ &0.224163 $\pm$ 0.000056(0.000056) 
              $\pm$0.000023(norm) \\
            & $ r_E^p[\rm fm] $ & {
              \boldmath $ 0.9022\pm 0.0016(0.0030) \pm 0.0021$ (norm)}\\ 
\hline
Polynomial     & $a_1$ & $-$0.2597 $\pm$ 0.0030(0.0030) $\pm$ 0.0560(norm) \\
$\chi^2/dof$   & $a_2$ &  0.0517$\pm$  0.0050( 0.0050) $\pm$ 0.0695(norm) \\
1.630          & $r_E^p[\rm fm]$ & 
\boldmath $ 0.8820 \pm 0.0051(0.0051) \pm 0.0951$ (norm)\\
\hline
\end{tabular}
\end{center}
\normalsize

\newpage

\noindent
We treat this fact as a signal that our estimators are slightly biased because
 all the fitted models are analytic with a true domain of analyticity and it has been 
expected that for all reliable fits the interval estimators of the model radii 
would belong to (or at least, well overlapping) the ``polynomial radius''
 interval estimator. 

Now, to show that possible biases are indeed presented in the above estimates, we 
repeat all the fits 
on the complete data sets only, but with the weights constructed without admixture
of the normalization errors as described below. 
The results are shown in \mbox{Table~5.3.}
They exhibit significant shifts in the adjustable parameters and  
radii values, however, one should hardly trust these estimates because 
of a bad quality of fits. The bad fits quality is a 
signal that either data from different experiments are inconsistent with each 
other in normalizations or the models under consideration have the limited 
areas of applicability.  So, we will try to construct a self-consistent data sample 
guided by the models and normalization errors in the data. 
 Then, we will try to find areas of application for each model.

 With each of the complete experimental data sets we associate the adjustable 
renormalization parameter  $1/\lambda^i$ to simulate a possible systematic 
rescaling of the data of different experiments 
with respect to each other.
These normalization parameters are a subject of evaluation by minimizing a 
least squares $\chi^2$ functional + ``penalty function''

{$$ \chi2 = \sum_{i=1}^{N}
{ 
 ({\vec y}^{\ i} - \lambda^{i} {\vec t}^{\ i}) 
\cdot W^{i} \cdot({\vec y}^{\ i} - \lambda^{i}{\vec 
t}^{\ i}) + 
 \sum_{i=1}^{N} \sum_{j=1}^{n^i}
\frac {(\lambda^{i} - 1)^2}{ {\sigma_i(j)}^2}  }, \eqno(2) $$}

\noindent where $i$ is the number of experiment,
$\lambda^i$ is the normalization for $i$'th experiment,
${\vec y}^{\ i}$ is the vector contained the experimental differential   
cross section values of the  $i$'th experiment,
${\vec t}^{\ i}$ is the vector contained the calculated cross-section values,
$W^{i}$ is the inverted error matrix for the data of the $i$'th experiment, 
$n^i$ is the number of points at the $i$'th set. 

$\lambda_i$ was calculated analytically at each step of minimization.
For details see Appendix 1. 

The second term in (2) is the ``penalty function'' 
increasing the $\chi^2$ value when the normalizations come out
from unity. 

Let us denote the total point-to-point normalization error quoted by 
experimentalists as $\nu_i(j)$,
where $j$ is the point number in the $i$'th
data set. Analogously, let $s_i(j)$ be the total systematic error
from all other sources combined in quadratures.\\

We apply the following procedure:
\begin{itemize}
\item
If systematic errors due to normalization were declared by authors,
we use $\sigma_i(j)=\nu_i(j)$;
\item
If $\nu_i(j)$ is not given, but $s_i(j)$ is given, then $\sigma_i(j) = s_i(j)$, with
   motivation that in this case we have a reference number and we adopt that
   $\nu_i(j) \le s_i(j)$. 
   In such a way we will obtain the upper limit of the normalization 
   systematic errors, since we cannot separate the non-normalization part 
   of error.

\item As has been indicated earlier we do not analyze the data which contained 
only total errors without any information about their sources. 
Certainly, we lose part of statistics ( about 7 \%), however, we will obtain more 
reliable results with full control of parameter errors.

\end{itemize}

So, at the end of the minimization we would have three sets of normalization 
factors, one for each model, and the same number of normalization
correlation matrices. 

The results of the fits with renormalizations are presented in Table~5.4.
We obtain the slightly better fits for each model but they all are still 
bad. We attribute this badness of fits either to the limited areas of the 
applicability of the models used or to the overestimation of the 
experimental errors in some experiments.

 Let us stress that now we have three
different estimates of the radius resulting from the polynomial fits, one for 
each set of the renormalization parameters. Now, we have the situation when the interval 
estimator of the radii derived from polynomial fits for each set of 
normalizations contains the corresponding interval estimator for the model radii
(as it should be when there are no rapid oscillations near $t = 0$).

As all the models give a rough equally
``good'' description, we cannot choose the ``best''  
model and the corresponding set of normalizations among others.
To proceed to the determination of the areas of applicability of the models 
selected we construct a common set of normalizations
following the procedure described in the Appendix 1 (``weighted averaging
of normalizations'', see Table~5.1).
 We will use this set of normalizations
 in the subsequent iterations of the fits. Their error 
matrix\footnote{Final arrays of normalizations and their error matrices
obtained for each model and their ``weighted averages'' can be
obtained by contacting polishchuk@mx.ihep.su.}~~  will be used in the calculations
of the parameter errors induced by our renormalization procedure.

 To check that we did not destroy drastically the previous results 
with three different sets of normalizations, we repeat all the fits with
obtained common renormalization parameters. 

The results are presented in Table~5.5.
As was expected we did get the ``slightly worse fits'' than those 
previously obtained with the individual sets of normalizations, but the whole 
picture of the results remains unchanged. All the three models give practically 
the same goodness of fit and closer to each other values of the proton 
electric radii.

\section{An attempt to get the confidential areas of applicability}

From previous sections we see that the ``naive'' (all errors quadratically 
combined) fits to the sample of complete data sets
 give  pretty good chisquares,
however, the fits to the same data sample with weights constructed from total 
errors without admixtures of correlated normalization errors result in 
unacceptably 
large chisquares. Using the renormalizations of the data admissable by 
the correlated normalization errors in the data we slightly improve the
 quality of the fits but they still remain unacceptable. This may be due to 
the following reasons or their combination:
\begin{itemize}
\item each model has own and restricted areas of applicability which 
should be determined;
\item experimental data have hidden systematics not quoted by experimentalists.
\end{itemize}

 One possibility to get  reliable parameter estimates is to use PDG scale 
factor prescription \cite{PDG} and to inflate parameter errors with 
scale-factor resulted from the fits (but not the errors due to the 
play with normalizations). But this way says nothing about the area of
models applicability and may be relevant only if the areas of applicability 
are known.

So, we have to determine the areas of applicability first or discover the 
subsample of data which has the overestimated experimental errors.
The way that can help to identify the areas of applicability of each
model is the following ``model driven filtration procedure.''\cite{Nikol1} 
\begin{itemize}

\item The whole physical region of kinematic variables $\log q^2$ versus 
$\cos \theta$\footnote{In these variables the distribution of points is
sufficiently uniform.}~~ is divided into bins and each parametrization is fitted to the data set
using normalizations obtained above. 

\item The value of $\chi^2/N_{points}$ is evaluated for each non-empty bin,
where $N_{points}$ is the number of points per bin. 
The bin is declared as belonging to ``the 
good description area'' (GDA) for a particular model if this value is less than
some threshold value. Otherwise, this bin is declared as belonging to
``the bad description area'' (BDA). The threshold normalized $\chi^2$ value is 
chosen in such a way that the confidence level (CL)
calculated for ``good description'' subset would exceed some threshold value
(say 50\% or 90\%). 
The procedure starts from sufficiently large $\kappa = {\chi^2/N}_{bin}=5$,
so at the first iteration all the points belong the GDA. Then, we decrease $\kappa$ 
by small steps, calculate CL at each value of the $\kappa$ and determine the
corresponding temporal GDA subsets until the threshold value is reached.

\item The ``good'' and ``bad'' data subsets are formed
from the bins classified in such a way. The ``good'' subset includes the points
belonging to GDA of all models and,
similarly, the ``bad'' subset contains the points belonging to BDA of all models.
\end{itemize}

The results of our standard fits of all models on the ``good at 90\% CL'' 
data sample are presented in Table~6.1. We see that rejection the ``bad''
data points not much affected on the values of the parameters for all models.
The interval estimators are slightly biased but the joint interval estimate
for the radius before filtration and after filtration are well overlapped.

Before the model driven filtration, we have
$$~~r^p_E \in [0.8912,0.9021] \quad [\rm fm] \quad {\rm three~models},$$
$$r^p_E \in [0.7818,0.9822] \quad [\rm fm] \quad {\rm polinomial}.$$

After the model driven filtration, we get
$$~~r^p_E \in [0.8955,0.9059] \quad [\rm fm] \quad {\rm three~models},$$
$$r^p_E \in [0.7847,0.9885] \quad [\rm fm] \quad {\rm polinomial}.$$

From these interval estimators (constructed using a linear combination of the 
experimental and normalization errors), we see that the polynomial estimator is 
inconclusive because it gives too wide interval estimator.

To show the stability of the radii under variation of the threshold CL value 
used in classifying data points as ``good'' and ``bad'', 
the radii obtained at two values of the CL = 50 \% and CL=90~\% are
tabulated below.
\small
\begin{center}
\begin{tabular}{||c|c|c|} 
\hline
Model       &     50\% CL,          &     90\% CL,           \\
            &    713 points         &    696 points          \\
            & $r_E^p$ in $[\rm fm]$ & $r_E^p$ in $[\rm fm]$  \\
\hline
Mack-98     & 0.8963 $\pm$ 0.0017   & 0.8937 $\pm$ 0.0015 \\
GK-98-2V    & 0.8976 $\pm$ 0.0011 & 0.8971 $\pm$ 0.0016 \\
GK-98-3V    & 0.9008 $\pm$ 0.0011 & 0.9005 $\pm$ 0.0010 \\
Polynomial  & 0.8842 $\pm$ 0.0058  & 0.8866 $\pm$ 0.0062  \\
\hline
\end{tabular}
\end{center}
\normalsize

\noindent
The quoted errors do not include the contributions due to normalization errors.

Now, let us discuss further the areas of applicability.  If the subset of bad 
data is not empty and is grouped in such a way that this 
subset can be interpreted as the area where all models failed then the 
good data subset belongs to the area of applicability with a preselected CL.

 If the bad data points are randomly distributed over the whole available 
kinematic region, 
then we have only one possibility:  To declare that the models are applicable 
in the whole region and the bad data are the data with overestimated errors 
(occasionally outstanding data). In this case, in order to estimate the errors in the
parameters, we should use the Birge scale-factor method.  
The situation with bad data bins is presented in Fig.3. 
There are no prominent
regularities in the population of bad data bins in the whole kinematic 
region and we conclude that our models are applicable in the whole kinematic 
region, except for a few bins with presumably overestimated experimental errors. 
The only way to get reliable errors is the usage of the scale factor 
prescription. As the best estimate of the proton electric radius in 
terms of average, dispersion and systematic errors, we take the estimate
derived from the GK-98-2V fit, taking into account the scale factor
$\sqrt{1.512}$ (see Table~5.5), and assign the additional systematic error
due to model selection as one half of the maximal difference between central 
values of the radii.
\noindent
So, we have
\vspace*{0.1cm}
$$ r_E^p = 0.900 \pm 0.002(exp) \pm 0.001(norm) \pm 0.002(models) \quad
[\rm fm].\vspace*{0.1cm}$$
The same procedure of constructing a final radius estimate based on the filtered
data (see Table~6.1) gives
\vspace*{-0.1cm}
$$ r_E^p = 0.897 \pm 0.002(exp) \pm 0.001(norm) \pm 0.003(models) \quad
[\rm fm].\vspace*{0.1cm}$$

We see that the estimate obtained with the model driven filtration gives the result
consistent with that obtained by using the Birge factor. So, our best 
estimate of the proton radius is the estimate based on the model driven 
filtration.

\vspace*{1cm}

\small
\begin{center}
\begin{tabular}{|c||c|c|} \hline
 Model name     & Free \& derived& Estimated parameter values and their errors\\
 \& Fit quality & parameters     &    \\   
\hline
Mack-98     & $x_1^s$       & 1.59 $\pm$ 0.17(0.13)$\pm$ 0.039(norm)\\
            & $x_1^v$       & 1.806 $\pm$ 0.020(0.017)$\pm$ 0.0056(norm)\\
            & $x_2^s$       & 0.0953 $\pm$ 0.0057(0.0045)$\pm$ 0.0018(norm)\\
            & $x_2^v$       & 0.0284 $\pm$ 0.0014(0.0010) $\pm$ 0.0004(norm)\\
$\chi^2/dof$& $C_1^{\rho}$  & $-$0.213 $\pm$ 0.011(0.011) $\pm$ 0.003(norm)\\
510.3/685   & $C_2^{\rho}$  & 0.6722 $\pm$ 0.0078(0.0077) $\pm$ 0.0019(norm)\\
 0.745      & $C_1^{\omega}$& $-$1.725 $\pm$ 0.020(0.020) $\pm$ 0.007(norm)\\
            & $C_2^{\omega}$& $-$9.07 $\pm$ 0.30(0.30) $\pm$ 0.09 (norm)\\
            &$\theta_1^{\omega}[\rm deg]$&91.521$\pm$0.014(0.014)
            $\pm$0.009(norm)\\
            &$\theta_2^{\omega}[\rm deg]$&91.414116$\pm$0.000021(0.000021)
            $\pm$0.000008(norm)\\
            &$\lambda[\rm GeV]$      &  0.0035 $\pm$ 0.0009(0.0007) 
            $\pm$ 0.0003(norm)\\      
            & $ r_E^p[\rm fm] $ & {
\boldmath $ 0.8937\pm 0.0015(0.0013) \pm 0.0008$ \bf (norm)} \\
\hline
GK-98-2V    & $\eta_1^v$      &0.005652 $\pm$0.000032(0.000032)
              $\pm$ 0.000021(norm)\\
            & $\eta_2^v$      &0.004407 $\pm$ 0.000077(0.000077)
              $\pm$ 0.000027(norm)\\ 
            & $\eta_1^s$      &0.03275 $\pm$ 0.00015(0.00015)
              $\pm$ 0.00011(norm)\\
            & $\eta_2^s$      &0.01253 $\pm$ 0.00055(0.00056)
              $\pm$ 0.00019(norm)\\
$\chi^2/dof$& $C_1^{\rho}$     &9.25 $\pm$ 0.19(0.15) $\pm$ 0.08(norm)\\
493.9/685   & $C_2^{\rho}$     &0.457 $\pm$ 0.010(0.010) $\pm$ 0.003(norm)\\
 0.721      & $C_1^{\omega}$  &$-$7.98 $\pm$ 0.20(0.17) $\pm$ 0.08(norm)\\
            & $C_2^{\omega}$  &1.594 $\pm$ 0.030(0.030) $\pm$ 0.011(norm)\\
            & $\theta_1^{\omega}[\rm deg]$& 91.50427 $\pm$ 0.00094(0.00095)
              $\pm$ 0.00068(norm)\\
            & $\theta_2^{\omega}[\rm deg]$& 91.415123 $\pm$ 0.000001(0.000001)
              $\pm$ 0.000001(norm)\\
            & $\lambda[\rm GeV]$ & 0.21991 $\pm$ 0.00023(0.00019) 
            $\pm$ 0.00008(norm)\\
            &$r_E^p[\rm fm]  $& {
\boldmath $  0.8971 \pm 0.0016(0.0014) \pm 0.0008$ \bf (norm) } \\
\hline
GK-98-3V    & $\eta_1^v$       &0.005844 $\pm$ 0.000032(0.000030)
            $\pm$ 0.000021(norm)\\
            & $\eta_2^v$ & 0.004495 $\pm$ 0.000073(0.000072)
            $\pm$ 0.000024(norm)\\
            & $\eta_1^s$       &0.03358 $\pm$ 0.00015(0.00014)
            $\pm$ 0.00010(norm)\\
            & $\eta_2^s$ & 0.01253 $\pm$ 0.00050(0.00050)
            $\pm$ 0.00016(norm)\\
$\chi^2/dof$& $C_1^{\rho}$   & 9.877 $\pm$ 0.077(0.072)
            $\pm$ 0.036(norm)\\
494.8/681   & $C_2^{\rho}$   & 0.4161 $\pm$ 0.0089(0.0084)
            $\pm$ 0.0028(norm)\\
 0.726      & $C_1^{\omega}$  & $-$8.798 $\pm$ 0.082(0.076)
            $\pm$ 0.036(norm)\\
            & $C_2^{\omega}$  & 1.688 $\pm$ 0.022(0.021) $\pm$ 0.009(norm)\\
            & $C_1^{\phi}$& 0.1937 $\pm$ 0.0053(0.0053)
            $\pm$ 0.0034(norm)\\
            & $C_2^{\phi}$& $-$0.137 $\pm$ 0.072(0.072) $\pm$ 0.020(norm)\\
            &$\theta_1^{\omega}[\rm deg]$& 91.50409 $\pm$ 0.00085(0.00085)
            $\pm$ 0.00062(norm)\\
            &$\theta_2^{\omega}[\rm deg]$& 91.415123 $\pm$ 0.000001 (0.000001)
            $\pm$ 0.000001(norm)\\
            &$\theta_1^{\phi}[\rm deg]$& 157.8  $\pm$ 2.6(2.6) 
            $\pm$ 1.9(norm)\\
            &$\theta_2^{\phi}[\rm deg]$& 159.9$\pm$ 13.0(13.0) 
            $\pm$ 8.0(norm)\\
            &$\lambda[\rm GeV] $ &0.219171 $\pm$ 0.000084(0.000085)
            $\pm$ 0.000038(norm)\\
            & $ r_E^p[\rm fm] $ & {
\boldmath $ 0.9005\pm 0.0010(0.0009) \pm 0.0006$ \bf (norm) } \\
\hline
Polynomial     & $a_1$ & $-$0.2624 $\pm$ 0.0036(0.0036) $\pm$ 0.0567(norm) \\
$\chi^2/dof$   & $a_2$ & 0.0575 $\pm$ 0.0059(0.0059) $\pm$ 0.0697(norm) \\
0.817          &$r_E^p[\rm fm]$ & { 
\boldmath $ 0.8866 \pm 0.0062(0.0062) \pm 0.0957(norm) $ }\\
\hline
\end{tabular}
\end{center}
\centerline{{\bf Table 6.1.} Results of the last iteration of fits on filtered data sample.}
\normalsize

\section{Summary}


We have made the comparison of the majority of published parametrizations of 
the proton electromagnetic FFs using the largest experimental data sample 
on $d\sigma/dt$ ever used.
All the models give unacceptable ``naive'' fit quality to the whole data sample 
and we have to 
select and modify three parametrizations for further detailed 
analyses. These three hybrid parametrizations give the best and close to each 
other values of the $\chi2/d.o.f.$ parameter which characterizes the fit 
quality (for the ``naive'' fits it is less than 1.2 for each model, see 
Table~4.1).

All the reconstructed models have correct analytical structures with the minimal 
set of known singularities. They are all easily generalized for inclusion of the
additional Wataghin terms \cite{wataghin},\cite{meshch94} into the GVMD 
multiplier of the FF models for
additional vector mesons as is seen from inspection formulas for the 
GK-98-2V and GK-98-3V parametrizations. It should be noted that the addition of 
$\phi$-meson did not resulted in improving the fits quality to the data in 
the space-like region (remember, that the addition of $\phi$-meson gives four 
extra adjustable parameters), but shifted 
slightly the proton radius towards higher values. 

The common feature of the tree models used is the presence of pionium 
contribution. In our preparatory attempts to
produce cross assessments with models without these contributions, we 
had to reject by all the filtrations more experimental data than in 
selected and modified models with pionium. In the case with pionium all the
filtrations rejected less than 18 \%. It is strong evidence in 
favour of the pionium contribution to the GVDM multiplier of the FFs.  
The  exact form of pionium contributions to the FFs is the task 
for the future refinements of our procedure and models.

It should be stressed that in the recent experiment 
conducted by L.L.Nemenov with co-workers at IHEP (Protvino) 
the pionium was detected experimentally \cite{nemenov}.


\vspace*{1.5cm}

\begin{figure}[H]
\centerline{\epsfig{file=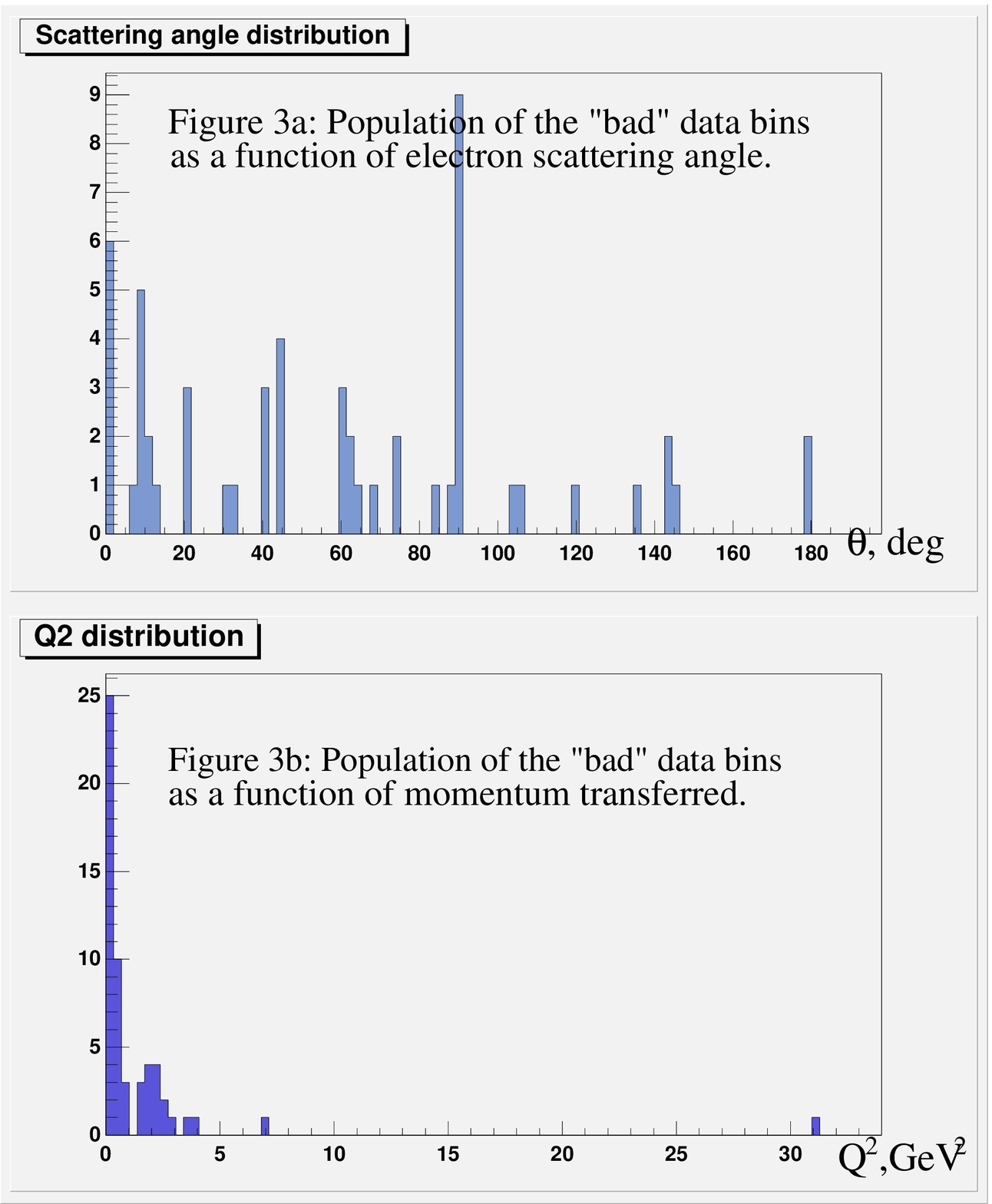,width=16cm}}
\caption{
Population of the bad data bins in the whole kinematic region.}
\end{figure}

\vspace*{0.5cm}

\begin{figure}[H]
\centerline{\epsfig{file=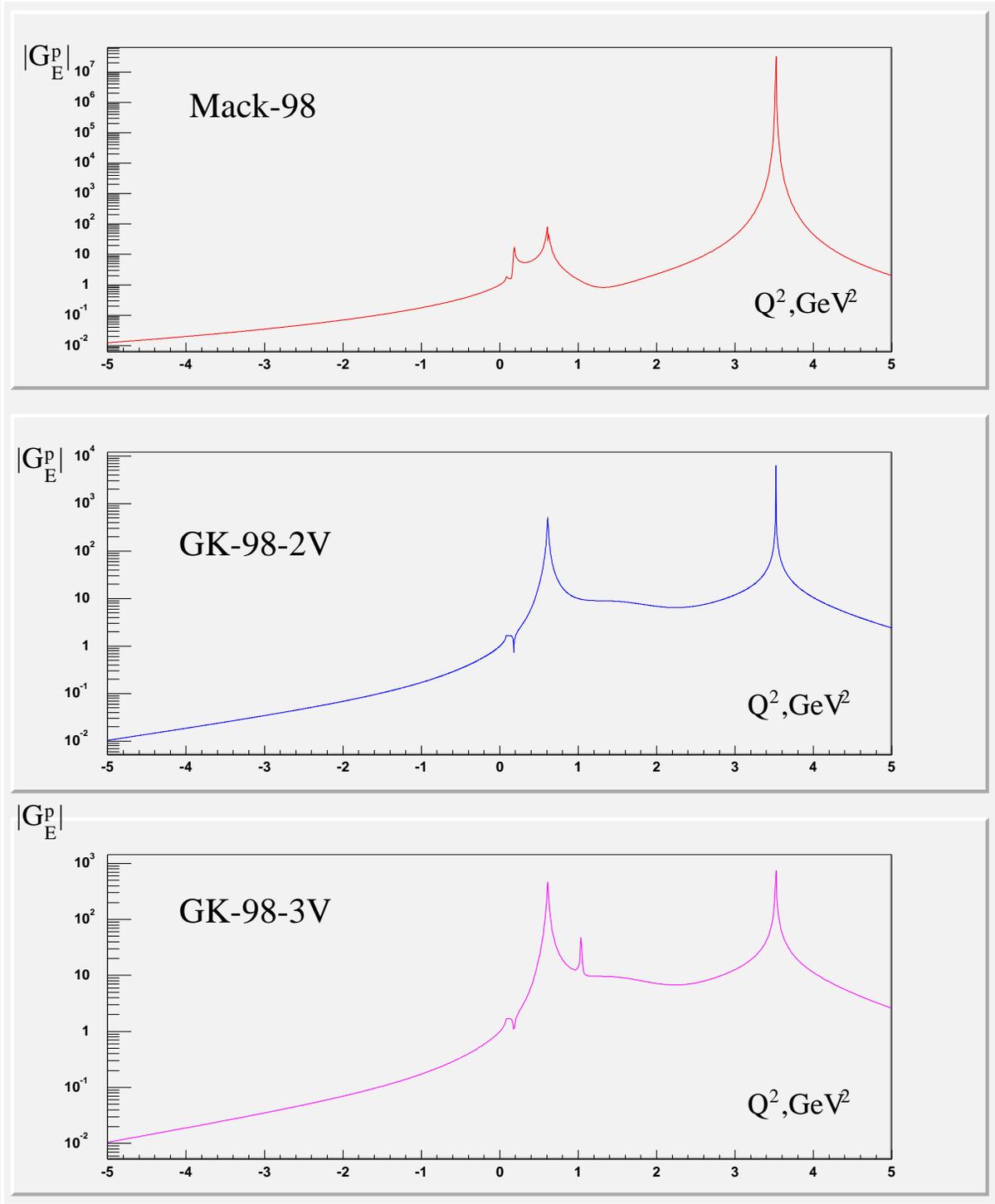,width=16cm}}
\caption{Modulus of the proton charge FFs given by three adjusted models.}
\end{figure}

\vspace*{0.5cm}

\begin{figure}[H]
\centerline{\epsfig{file=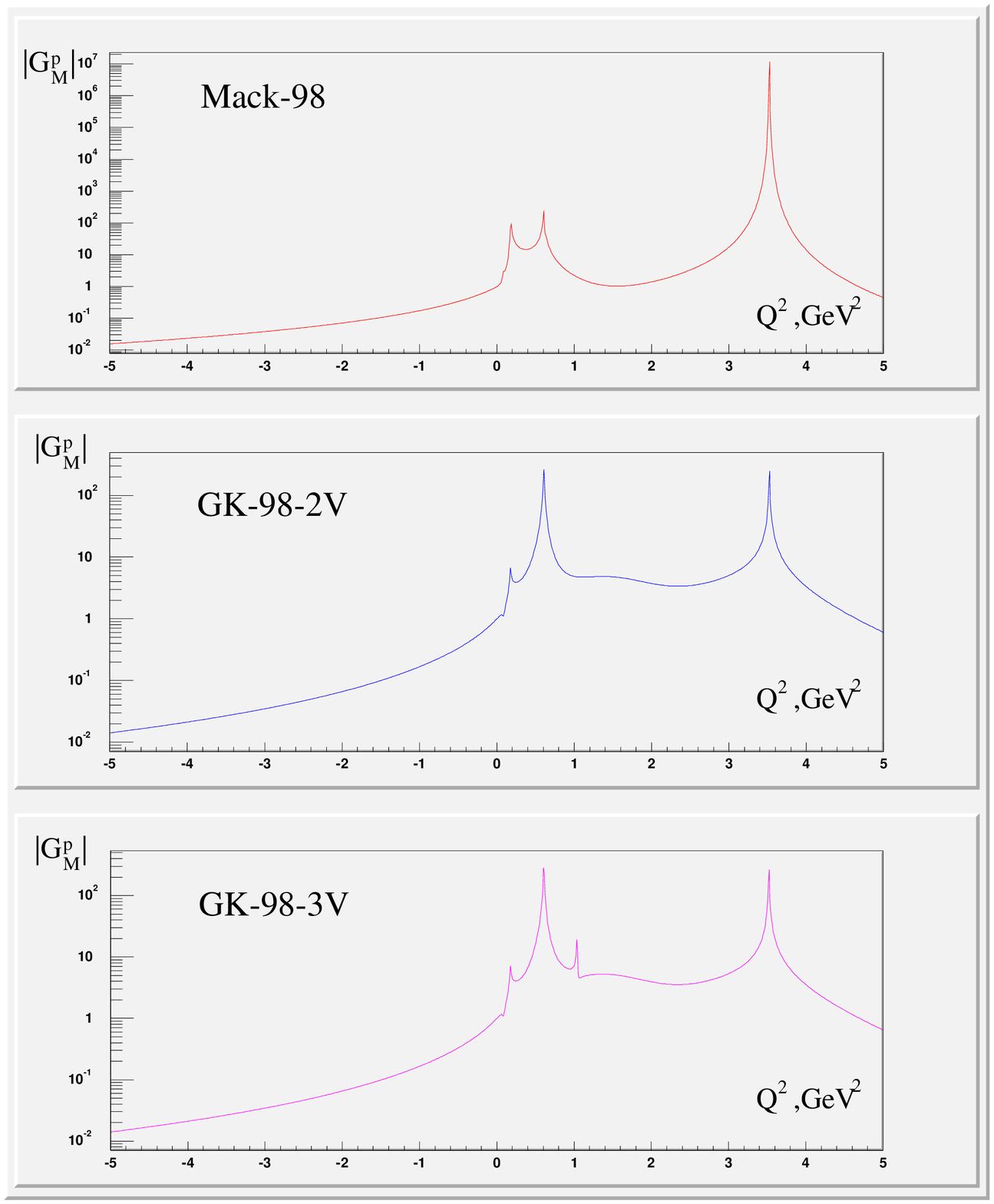,width=16cm}}
\caption{Modulus of the neutron magnetic FFs given by three adjusted models.}

\end{figure}

\newpage

As it is seen from Tables~5.5 and~6.1 all three models give
statistically equivalent descriptions of the unfiltered and filtered data. 
It can be also seen in Figs. 4 - 5, where the modulae of the electric 
and magnetic formfactors
for each model are presented for comparison in a large interval of
negative and positive $q^2$.

The model GK-98-2V is preferable at this stage of data and models cross 
assessment because it has a proper behaviour at high $q^2$ and gives the best 
description of the assessed data sample.
The main results are summarized as follows:
\begin{itemize}
\item We have developed {\it the data and models cross assessments (CRAS) 
technology} in the case of the data on elastic $e^{\pm} p$ scattering  and 
models of the nucleon form factors. We believe that the CRAS technology will 
be applicable in other areas as well and will be helpful in the following
tasks:
\begin{itemize}
\item to make  data from different experiments mutually consistent on the 
basis of available relevant models;
\item to determine the areas of applicability of the competitive models;
\item to produce technical and derived physical parameters estimations with
metrological quality.
\end{itemize}
\item Using the CRAS technology we have constructed, from the well known inputs,
 the currently best parametrization of the nucleon FFs in the space-like region 
which is extendable in the sense of GVDM,
can be also used to describe nucleon FFs in the time-like region, and is 
capable to eliminate the ``false'' contradiction between the charge radius 
of the proton from the hydrogen Lamb shift measurements (see recent review of
S.G.~Karshenboim \cite{karsh97}) and from the high-energy electron-proton 
elastic scattering data.  
\end{itemize}

From the fit results of all the models to the ``90\%{\quad}{\rm CL}'' sample, 
we have derived our best estimate of the proton charge radius

$$ r_E^p = 0.897 \pm 0.002(exp) \pm 0.001(norm) \pm 0.003(models) \quad 
[\rm fm]$$ 
given by the modified Gary-Kr\"umpelman model on the filtered data sample. This
estimate is in good agreement with the estimate derived from the fits to 
unfiltered data but with the usage of the Birge method to justify the errors assigned 
to the adjustable parameters
$$ r_E^p = 0.900 \pm 0.002(exp) \pm 0.001(norm) \pm 0.002(models) \quad 
[\rm fm].$$
The traditional method of extracting proton radius from quadratic polynomial
fit gives the inconclusive estimates on the model driven renormalized data 
(see Table~5.5 and Table~6.1), because they have too large systematic errors
due to more correct treatment of the systematic error in the radius induced
by the correlated experimental errors of normalization. 

It should be noted that the above results on the parameters of the three
models analysed are still preliminary. They are a subject for the re-estimation
with the data on $n~e^{-}$, $e^{\pm}~n$ elastic scattering and 
$e^+ e^- \to$~{\it nucleon~anti-nucleon} added to the 
data on $e^{\pm}~p$ elastic scattering. It is likely that all models will
require the inclusion of the protonium atom singularities. We plan to perform
further CRAS-iterations for the nucleon form factors with an extended data 
sample in the near future. 

A straightforward outcome of our analyses for the current and planned 
experiments are as follows: 
\begin{itemize}
\item The HERA data on $e^{\pm} p \to e^{\pm} p$ would be extremely
useful not only for the extraction of the FFs but also to check the relevancy of one
photon approximation in the description of the $d\sigma/dt$ data at moderate
 $-t$. Unfortunately, these measurements are not discussed or planned in the
HERA physics program.
\item The CEBAF data near $t = 0$ will be very helpful to test the GVDM with
pionium contribution (note that ``bad'' data points are concentrated 
near $t = 0$, see Fig.3.)   
\end{itemize}

To simplify the reproduction of our results, we prepare a 
special ``WEB $e^{\pm} p \to e^{\pm} p$ data site'' with  
computer readable files of all data, models, and current results for the
arrays of parameter values and their correlators on the basis of PPDS data.
It can be reached by

\begin{verbatim}
               http://sirius.ihep.su/~polishch/ep2ep_tot.html 
\end{verbatim}
or
\begin{verbatim}
               http://dbserv.ihep.su/~bvp/ep2ep_tot.html 
\end{verbatim}

\section{Acknowledgements}
We are thankful to S.R.~Slabospitsky for his careful study of the 
manuscript and suggestions to make it more readable.
This work was supported in part by Russian Foundation for Basic 
Research by Grants: RFBR-93-02-3733, RFBR-96-07-89230, and 
RFBR-99-07-90356.

\newpage

\appendix

\section{Normalizations and their correlations}
\noindent
We shall use the following notation:
\begin{description}
\item[\Large {$  {\vec y}^{\ i} $}] ---  array of average values of an 
observable $y(x)$ measured at $n^i$ x-points in $i$-th experiment.
\item[ \Large  {$ T^{i}$}] --- $n^i \times n^i$ error matrix of $i$-th experiment.
\item[ \Large {$ W^{i} = (T^i)^{-1}$}]  --- weight matrix for
the data of $i$-th experiment.
\item[ \Large {$  \nu_{i}(j) $}] --- estimation of normalization 
uncertainty in $j$-th point of the $i$-th experiment.
\item[ \Large {$ \lambda^{i}$}] --- auxiliary normalization factor to 
be assigned to $i$-th experiment to tune mutual normalizations of different
independent experiments.
\end{description}

We shall use the following expression for {$ \chi2$}
{$$ \chi2 = \sum_{i=1}^{N}
{ 
 ({\vec y}^{\ i} - \lambda^{i} {\vec t}^{\ i}) 
\cdot W^{i} \cdot({\vec y}^{\ i} - \lambda^{i}{\vec 
t}^{\ i}) + 
 \sum_{i=1}^{N} \sum_{j=1}^{n^i}
\frac {(\lambda^{i} - 1)^2}{ {\nu_i(j)}^2}  }, \eqno(1) $$}

\noindent
which must be minimized with respect to model parameters $\vec p^{\ m}$ and 
with respect to auxiliary normalization factors.
Necessary requirements for extremums lead to equations

{$$ \frac{\partial \chi2^{m}}{\partial p^m_{\alpha_m}} = 
-2 \sum_{i=1}^{N}{\lambda^{m,i} \frac{\partial \vec t^{\ m}(i)}
{\partial p^{m}_{\alpha_m}} 
\cdot W^{i} ({\vec y}^{\ i} - \lambda^{m,i}{\vec t}^{\ m}(i))} = 0, \eqno(2)$$}

{$$ \frac{\partial \chi2^{m}}{\partial \lambda^{m,i}} = 
-2{\vec t}^{\ m}(i) \cdot 
W^{i} ({\vec y}^{\ i} - \lambda^{m,i}{\vec t}^{\ m}(i)) + 
2 \sum_{j=1}^{n^i} \frac {\lambda^{m,i}-1}{ {\nu_i(j)}^2 } = 0. \eqno(3) $$}

\noindent
From the second set of equations one can express auxiliary normalizations 
for $m$-th model as follows:

 {$$ \lambda^{m,i} = 
\frac{{\vec t}^{\ m}(i) \cdot W^{i}{\vec y}^{\ i} + \sum_{j=1}^{n^i} 1/{\nu_{i}(j)}^2}
{{\vec t}^{\ m}(i) \cdot W^{i}{\vec t}^{\ m}(i) + \sum_{j=1}^{n^i} 1/{\nu_{i}(j)}^2}~. \eqno(4) $$}

\noindent
Inserting these expressions for normalization factors into above expression for 
{$ \chi2^{m}$}, we will get a new formula
for {$ \chi2^{m}$}, which must be minimized with respect to model 
parameters only. 
After solving the minimization task and selecting proper local minimum 
point {$ \langle p_{\alpha_m}^{m} \rangle $}, we can obtain 
the corresponding error matrix
{$$ \Vert \frac{1}{2}  \frac{\partial \chi2^{m}}
{\partial p_{\alpha_m}^{m} \partial p_{\beta_m}^{m}} \Vert^{-1} \ 
= \langle \Delta p_{\alpha_m}^{m} \Delta p_{\beta_m}^{m} \rangle \ = \ 
\Pi^{m}_{\alpha_m \beta_m}. $$}

\noindent
In a general case the normalization parameters for different models 
are correlated,
and some of the normalizations may be strongly correlated for one model.
Let us first prepare formulae for correlators
between normalizations for different models.

$$
N^{mm'}_{ij} = 
\langle (\sum_{\alpha_m} 
{\frac {\partial \lambda^{m,i}} {\partial p^{m}_{\alpha_m}}}
\Delta p^m_{\alpha_m} +
\sum_{\mu_i}{\frac {\partial \lambda^{m,i}}{\partial y^{i}_{\mu_i}}}
\Delta y^i_{\mu_i})
(\sum_{\beta_{m'}} 
{\frac {\partial \lambda^{m',j}} {\partial p^{m'}_{\beta_{m'}}}}
\Delta p^{m'}_{\beta_{m'}} +
\sum_{\mu_j}{\frac {\partial \lambda^{m',j}}{\partial y^{j}_{\mu_j}}}
\Delta y^j_{\mu_j}) \rangle $$

$$ = \sum_{\alpha_m \beta_{m'}} 
{\frac {\partial \lambda^{m,i}} {\partial p^{m}_{\alpha_m}}}
{\frac {\partial \lambda^{m',j}}{\partial p^{m'}_{\beta_{m'}}}}
\Pi^{mm'}_{\alpha_m \beta_{m'}} 
+ \sum_{\beta_{m'} \mu_i}
{\frac {\partial \lambda^{m,i}} {\partial y^{i}_{\mu_i}}}
{\frac {\partial \lambda^{m',j}} {\partial p^{m'}_{\beta_{m'}}}}
{\frac {\partial p^{m'}_{\beta_{m'}}} {\partial y^{i}_{\mu_i}}}
(\sigma^i_{\mu_i})^2 $$
$$ \qquad + \sum_{\alpha_m \mu_j}
{\frac {\partial \lambda^{m,i}} {\partial p^{m}_{\alpha_m}}}
{\frac {\partial \lambda^{m',j}} {\partial y^{j}_{\mu_j}}}
{\frac {\partial p^{m}_{\alpha_m}} {\partial y^{j}_{\mu_j}}}
(\sigma^j_{\mu_j})^2 
+ \delta_{ij} \cdot \sum_{\mu_i} 
{\frac {\partial \lambda^{m,i}} {\partial y^i_{\mu_i}}}
{\frac {\partial \lambda^{m',i}} {\partial y^i_{\mu_i}}}
(\sigma^i_{\mu_i})^2 ,  \eqno(8) $$

\noindent
where

$$\Pi^{mm'}_{{\alpha_m}{\beta_{m'}}} = \frac{1}{4} \sum_{\eta,\nu}\Pi^{m}_{\alpha_m \eta}
  \Pi^{m'}_{\beta_{m'} \nu} \sum_{i,\mu_i}
\frac {\partial^2 \chi2^m}{\partial p^m_{\eta} \partial y^i_{\mu_i}}
\frac {\partial^2 \chi2^{m'}}{\partial p^{m'}_{\nu} \partial y^i_{\mu_i}}
(\sigma^i_{\mu_i})^2, $$

\noindent
and partial derivatives are estimated at minima points for $m$ and $m'$
models.
Now we are able to construct estimators for a unique set of lambdas and
their error matrix

$$ \langle \lambda^i \rangle = (\sum_m \frac{1}{N^{mm}_{ii}})^{-1} 
\sum_m \frac{\lambda^{m,i}}{N^{mm}_{ii}}
\eqno(9) $$

\noindent
and

$$N_{ij} = (\sum_m \frac{1}{N^{mm}_{ii}})^{-1}
(\sum_m \frac{1}{N^{mm}_{jj}})^{-1}
\sum_{mm'} \frac {N^{mm'}_{ij}}
{N^{mm}_{ii}N^{m'm'}_{jj}}~. \eqno(10) $$ \\

\noindent
\section{Error propagation}

\subsection{Errors of model parameters}

\noindent
Having numerical values for normalization correction factors, we can find 
a final parameter estimation. We will use the standard expression for modified
{$ \chi2^{m}$}

{$$ \chi2^{m} = \sum_{i=1}^{N}
{ 
({\vec y}^{\ i} - \lambda^{m,i} {\vec t}^{\ m}(i) ) 
\cdot W^{i}({\vec y}^{\ i} - \lambda^{m,i}{\vec 
t}^{\ m}(i))  },$$}

\noindent
with axillary correction factors fixed at their best values obtained
above. After minimization we will obtain new estimations for model parameters 
and corresponding parameter errors

{$$ \langle  p_{\alpha_m}^{m} \rangle, \  
\langle \Delta p_{\alpha_m}^{m} \Delta p_{\beta_m}^{m} \rangle^{expt} = 
\Pi^m_{\alpha_m \beta_m}. $$}
\noindent
To estimate the contribution to parameters covariance from renormalization,
we will propagate covariances $N_{ij}^{m}$ as follows:

{$$ \Pi^{m,norm}_{\alpha_m \beta_m} \ = \ 
\langle \Delta p^m_{\alpha_m} \Delta p^m_{\beta_m} \rangle^{norm} \ = \
\sum_{ij}^N {\frac{\partial p^m_{\alpha_m}}{\partial \lambda^i} N^m_{ij}
\frac{\partial p^m_{\beta_m}} {\partial \lambda^j} }. $$} \\
\noindent
Following the same way, we can propagate systematic (or statistical)
point-to-point errors to be the errors of parameters:

{$$ \Pi^{m,sys}_{\alpha_m \beta_m} \ = \ 
\langle \Delta p^m_{\alpha_m} \Delta p^m_{\beta_m} \rangle^{sys} \ = \
\sum {\frac{\partial p^m_{\alpha_m}}{\partial y^i_{\mu_i}} N_{ij}^{\mu_i \nu_j}
\frac{\partial p^m_{\beta_m}} {\partial y^j_{\nu_j}} }, $$} \\
\noindent
here $\mu_i$ and $\nu_j$ are the $\mu$'th point in the $i$'th experiment
and $\nu$'th point in the $j$'th experiment, respectively. 
If usual MINUIT assumptions about the minimum are valid, 
this matrix must coincide with the MINUIT error matrix.
Really, we have at the minimum position

{$$\Delta \chi^2 = \frac{1}{2} \frac{\partial \chi2} 
{\partial p_{\alpha} \partial p_{\beta}}
{{\langle \Delta p_{\alpha} \Delta p_{\beta} \rangle}^{calc}}, 
$$} \\
in the absence of point-to-point correlation
{$$
{{\langle \Delta p_{\alpha} \Delta p_{\beta} \rangle}^{calc}} =
\sum {\frac{\partial p_{\alpha}}{\partial y^i_{\mu_i}}
\frac{\partial p_{\beta}} {\partial y^i_{\mu_i}} } {(\sigma^{\mu_i}_{i})}^2,
$$} \\
{$$\Delta \chi^2/n_{points} = \frac{1}{2} \frac{\partial \chi2}
{\partial p_{\alpha} \partial p_{\beta}}
\frac{\partial p_{\alpha}}{\partial y^i_{\mu_i}}
\frac{\partial p_{\beta}} {\partial y^i_{\mu_i}}  {(\sigma^{\mu_i}_{i})}^2 =
\frac{1}{2} \frac{\partial \chi2} {\partial y^i_{\mu_i} \partial y^i_{\mu_i}  } = 1, 
$$} \\
it corresponds to the standard MINUIT error definition. \\

To calculate partial derivatives of parameters over lambdas, we will use 
    the usual trick\footnote{We thank our colleague S.I.~Alekhin who point out
    this method} \cite{LB}.
    The set of equations 
 
{$$\frac{\partial \chi2^m}{\partial p^m_{\alpha_m}} = 0$$}

\noindent might be considered as the set of functional equations defining dependencies of
parameters upon lambdas. Taking the total derivative from both parts of above
equations over $\lambda^{m,i}$, we will obtain a linear system

{$$\sum_{\beta_m} ({\frac {\partial^2 \chi 2^m}{\partial p^m_{\alpha_m} 
\partial p^m_{\beta_m}} }
{\frac {\partial p^m_{\beta_m}}{\partial \lambda^{m,i} }} ) + 
{\frac {\partial^2 \chi 2^m}{\partial p^m_{\alpha_m} \partial \lambda^{m,i}} } \ 
= \ 0, $$} \\
\noindent
from which

{$$ \frac {\partial p^m_{\alpha_m}}{\partial \lambda^{m,i}} \ = -\
\sum_{\beta} \frac{1}{2} \langle \Delta p^m_{\alpha_m} \Delta p^m_{\beta_m} 
\rangle^{expt} 
{\frac {\partial^2 \chi 2^m}{\partial p^m_{\beta_m} \partial \lambda^{m,i}}}~. $$}
\noindent
Finally, we will have

{$$ \Pi^{m,norm}_{\alpha_m \beta_m} \ = \
\frac{1}{4} \sum_{ij}^N \sum_{\mu \nu} 
\Pi^m_{\alpha_m \mu} 
{\frac {\partial^2 \chi 2^m}{\partial \lambda^{m,i} \partial p^m_{\mu}} } 
N^m_{ij}
{\frac {\partial^2 \chi 2^m}{\partial \lambda^{m,j} \partial p^m_{\nu}} }
\Pi^m_{\alpha_m \nu} $$
\noindent
with 
{$$ \frac {\partial^2 \chi 2^m}{\partial \lambda^{m,i} \partial 
p^m_{\mu} } \ = \ 
2 \frac {\partial {\vec t}^{\ m}(i)}{\partial p^m_{\mu}} \cdot W^{i} (2 
\lambda^{m,i} {\vec t}^{\ m}(i) - {\vec y}^{\ i}). $$}
\noindent
Thus, for each model $t^{m}$ we will have full estimations
of parameters and their covariances
{
$$ \langle p_{\alpha}^{m} \rangle, \ \Pi^{m}_{\alpha_m \beta_m},\ 
 \Pi^{m,norm}_{\alpha_m \beta_m}. $$}

\subsection{Errors of proton charge radius}

\noindent
Errors of the proton charge radius are calculated by propagation of the
 parameter errors of each type separately.
\noindent
 The ``systematic'' error due to the 
uncertainties in normalizations

{$$ \Delta^{2}_{m,(norm)} \ = \ 
\sum_{\alpha_m \beta_m} {\frac{\partial r_E^{p,m} }{\partial p^m_{\alpha_m}}
 \Pi^{m,norm}_{\alpha_m \beta_m}
\frac{\partial r_E^{p,m} } {\partial p^m_{\beta_m}} }~. $$} \\
\noindent
The error due to the ``total experimental'' parameters errors obtained from 
MINUIT system 

{$$ \Delta^{2}_{m} \ = \ 
\sum_{\alpha_m \beta_m} {\frac{\partial r_E^{p,m} }{\partial p^m_{\alpha_m}}
 \Pi^{m}_{\alpha_m \beta_m}
\frac{\partial r_E^{p,m} } {\partial p^m_{\beta_m}} }~.$$} \\

\end{document}